\documentclass[a4paper]{article}
\pdfoutput=1
\usepackage{jheppub}

\usepackage[utf8]{inputenc}

\usepackage{amsmath}

\usepackage{amssymb}
\usepackage{graphicx,color}

\usepackage{amsmath,amsthm}

\usepackage{mathrsfs}

\usepackage{bm}

\newcommand{\Ref}[1]{(\ref{#1})}

\newtheorem{Theorem}{Theorem}[section]

\newtheorem{Lemma}[Theorem]{Lemma}

\newcommand{\half}{\frac{1}{2}}

\newcommand{\ccirc}{\kern0.2ex\vcenter{\hbox{$\scriptstyle\circ$}}\kern0.2ex}

\def\be{\begin{eqnarray}}
\def\ee{\end{eqnarray}}

\newcommand{\ce}{\mathcal E}

\newcommand{\cp}{\mathcal P}

\newcommand{\calr}{\mathcal R}
\newcommand{\cs}{\mathcal S}

\newcommand{\fa}{\mathfrak{a}}

\newcommand{\fh}{\mathfrak{h}}

\newcommand{\fs}{\mathfrak{s}}

\renewcommand{\a}{\alpha}
\renewcommand{\b}{\beta}
\newcommand{\g}{\gamma}

\newcommand{\eps}{\varepsilon}

\newcommand{\sig}{\sigma}

\renewcommand{\l}{\lambda}
\renewcommand{\L }{\Lambda}

\renewcommand{\t}{\tau}

\newcommand{\rmd}{\mathrm d}

\newcommand{\lt}{\left}
\newcommand{\rt}{\right}

\newcommand{\tr}{\mathrm{tr}}

\newcommand{\Ar}{\mathrm{Ar}}

\newcommand{\sgn}{\mathrm{sgn}}

\title{Improved ($\bar{\mu}$ - Scheme) Effective Dynamics of Full Loop Quantum Gravity}

\author[1,2]{Muxin Han}  

\affiliation[1]{Department of Physics, Florida Atlantic University, 777 Glades Road, Boca Raton, FL 33431-0991, USA}

\affiliation[2]{Institut f\"ur Quantengravitation, Universit\"at Erlangen-N\"urnberg, Staudtstr. 7/B2, 91058 Erlangen, Germany}

\author[3,4]{\ Hongguang Liu}  

\affiliation[3]{Aix Marseille Univ, Université de Toulon, CNRS, CPT, Marseille, France}

\affiliation[4]{Center for Quantum Computing, Pengcheng Laboratory, Shenzhen 518066, China}

\emailAdd{hanm(At)fau.edu}
\emailAdd{liu.hongguang(At)cpt.univ-mrs.fr}

\date{\small\today}
 
\abstract{We propose a new derivation from the full Loop Quantum Gravity (LQG) to the Loop Quantum Cosmology (LQC) improved $\bar{\mu}$-scheme effective dynamics, based on the reduced phase space formulation of LQG and a proposal of effective Hamiltonian/action in the full LQG. A key step of our program is an improved regularization of the full LQG Hamiltonian on a cubic lattice. The improved Hamiltonian uses a set of ``dressed holonomies'' $h_\Delta(\fs)$ which not only depend on the connection $A$ but also depend on the length of the curve $\fs$. With the improved Hamiltonian, we propose a quantum effective action and derive a new set effective equations of motion (EOMs) for the full LQG. Then we show that these new EOMs imply the $\bar{\mu}$-scheme effective dynamics for both the homogeneous-isotropic and Bianchi-I cosmology, and predict bounce and Planckian critical density. As a byproduct, although the model is defined on a cubic lattice, we find that the improved effective Hamiltonian of cosmology is invariant under the lattice refinement. The cosmological effective dynamics, predictions of bounce and critical density are results at the continuum limit.
}

\keywords{}

\begin{document}

\maketitle

\section{Introduction}

Loop Quantum Gravity (LQG) is a promising attempt toward a non-perturbative and background independent theory of quantum gravity (see e.g. \cite{book, review} for reviews). Among many important achievements of LQG, one of the most profound physical predictions is the resolution of singularity e.g. \cite{Bojowald:2001xe,Ashtekar:2006wn,Singh:2009mz,Assanioussi:2019iye,Ashtekar:2018cay,Ashtekar:2018lag,Assanioussi:2019twp,Gambini:2013hna,BenAchour:2018khr,Rovelli:2014cta,Han:2016fgh,Han:2019vpw}. It is well-known that the classical theory of Einstein gravity breaks down at singularities, while the purpose of quantum gravity is to extend the gravity theory to describe the physics of singularities. 

A well-developed theme of singularity resolution is Loop Quantum Cosmology (LQC) where the big-bang singularity is replaced by a quantum bounce (see e.g. \cite{Bojowald:2006da,Ashtekar:2008zu,Agullo:2016tjh} for reviews). LQC applies the LQG method to the homogeneous and isotropic sector of gravity. The homogeneous and isotropic sector is given by a classical symmetry reduction from infinitely many degrees of freedom (DOFs) of gravity to a single DOF (the scale factor). The quantum dynamics of LQC have been studied extensively, and turns out that it can be efficiently described by an effective equation, which reduces to classical Friedmann equation at low energy density, while modifying Friedmann equation at high energy density \cite{Taveras:2008ke}. The solution of effective equation demonstrates that the big-bang singularity is resolved and replaced by a big bounce, where the curvature is finite and Planckian. The time evolution of cosmology governed by the LQC effective equation is often called the effective dynamics.  

Due to the theme of symmetry-reduction before quantization, LQC has been suffered from the long-standing issue on the relation with the full theory of LQG. Symmetry-reduced models of loop quantum black holes share the same issues. There have been interesting recent progresses toward this relation \cite{Han:2019vpw, Alesci:2013xd, Bodendorfer:2014vea, Bodendorfer:2015hwl, Alesci:2016rmn, Dapor:2017rwv, Engle:2007zz, 2016arXiv160105531H, Fleischhack:2010zt, Rovelli:2008aa,Calcagni:2014tga}. Although there is a top-down derivation from the full LQG to LQC $\mu_0$-scheme effective dynamics \cite{Han:2019vpw}, the LQG derivation to the improved $\bar{\mu}$-scheme effective dynamics of LQC is still largely open \cite{Dapor:2019mil}. The $\bar{\mu}$-scheme is physically preferred because it predicts a constant Planckian energy density at the bounce, while the density at the bounce in the $\mu_0$-scheme can vary and possibly non-Planckian.

Our present work makes one step further toward resolving the above issue, and proposes a new derivation of the LQC improved $\bar{\mu}$-scheme effective dynamics from the full LQG. Our derivation is based on the reduced phase space formulation of LQG \cite{Giesel:2007wi,Giesel:2007wn}. A key step of our program is an improved regularization of the full LQG Hamiltonian on a cubic lattice. The improved Hamiltonian uses a set of ``dressed holonomies'' $h_\Delta(\fs)$ which not only depend on the connection $A$ but also depend on the length of the curve $\fs$. With the improved Hamiltonian, we propose a quantum effective action and derive a new set effective equations of motion (EOMs) for the full LQG. Then we show that these new EOMs imply the $\bar{\mu}$-scheme effective dynamics for both the homogeneous-isotropic and Bianchi-I cosmology, and predict bounce and Planckian critical density. Moreover, although the model is defined on a cubic lattice, we find that the improved effective Hamiltonian of cosmology is invariant under the lattice refinement, and has a trivial continuum limit. The predictions of bounce and critical density are also independent of the lattice refinement.

The idea of constructing the improved Hamiltonian ${\bf H}_\Delta$ is to regularize the curvature $F(A)$ of the Ashtekar-Barbero connection $A$ in a non-conventional manner. In the standard regularization \cite{QSD}, $F(A)$ is replaced by a loop holonomy, which is a functional of $A$ only. In our approach, we denote by $S(\Delta)$ the surface enclosed by the loop holonomy $h(\Delta)$, and require $h(\Delta)$ to be not only a functional of $A$ but also depending on the geometry of the surface $S(\Delta)$. Inspired by LQC, we introduce an area scale $\Delta$ such that the area of $S(\Delta)$ is fixed to be $\Delta$. Here $\Delta$ is a free parameter of dimension (length)$^2$ and may be chosen to be the minimal area gap in LQG. $h(\Delta)$ depends on $\Delta$, so does the Hamiltonian ${\bf H}_\Delta$ constructed by $h(\Delta)$. Our regularization of ${\bf H}_\Delta$ is described in Section \ref{Improved Hamiltonian}.

Interestingly, ${\bf H}_\Delta$ is a \emph{non-graph-changing Hamiltonian which changes the graph}. ${\bf H}_\Delta$ is non-graph-changing because it is a function on the phase space $\cp_\g$ of holonomies and fluxes on a fixed cubic lattice $\g$. But it changes the graph $\g$ because $h(\Delta)$ contains holonomies along curves which do not belong to $\g$. The simple way to make these two aspects of ${\bf H}_\Delta$ consistent is to define $h(\Delta)$ as a phase space function on $\cp_\g$. Indeed a desired $h(\Delta)$ can be defined by comparing the continuum approximations of $h(\Delta)$ and the holonomy around a plaquette in $\g$. The procedure involves certain gauge fixing. The strategy of constructing $h(\Delta)$ is discussed in Sections \ref{Holonomy along Planckian Segment} and \ref{Loop Holonomy around Planckian Plaquette}.

In Section \ref{Path Integral and Effective Equations}, we propose a canonical quantum effective action $S_{eff}$ with the improved Hamiltonian ${\bf H}_{\Delta}$. ${\bf H}_{\Delta}$ and $S_{eff}$ explicitly depends on the scale $\Delta$ which is similar to UV cut-offs in quantum effective actions of quantum field theories. It suggests that $S_{eff}$ should be viewed as a quantum effective action which takes into account quantum effects. We derive a new set of EOMs of the full LQG from the variational principle of $S_{eff}$. These EOMs are \emph{improved effective equations} because they come from the improved Hamiltonian and relate to the $\bar{\mu}$-scheme effective dynamics in LQC.


In Section \ref{Homogeneous Effective Dynamics}, we look for spatial homogeneous solutions of our improved effective equations of LQG. When inserting ansatz that respects the spatial homogeneous symmetry, we show that the effective equation reduces to $\bar{\mu}$-scheme effective equations of Bianchi-I LQC. If we further restrict the solution to be isotropic, the effective equation reduces to $\bar{\mu}$-scheme effective equations of the standard LQC. We demonstrate the singularity resolution and bounce in Section \ref{Cosmic Bounce}, and reproduce the critical density $\rho_c$ at the bounce to be constant $\rho_c=\frac{16-\beta^2  \Delta  \Lambda }{\beta^2  \Delta  \kappa }$. $\rho_c$ is Planckian and corresponds to the Planckian curvature when $\Delta\sim\ell_P^2$. Moreover although ${\bf H}_\Delta$ of the full theory involves the gauge fixing, the cosmological effective dynamics is gauge invariant and independent of the gauge fixing.

Additionally in Section \ref{Lattice Independence}, we observe that although ${\bf H}_\Delta$ is defined on the lattice $\g$, it is invariant under lattice refinement therefore has a trivial continuum limit, at least when evaluating at homogeneous solutions. Effective equations, predictions of bounce and critical density are also invariant under the lattice refinement, so can be understood as results at the continuum limit. The key point here is that $h(\Delta)$ is defined around a surface with fixed area and is invariant under the lattice refinement. This lattice independence of ${\bf H}_\Delta$ suggests that the theory at the homogeneous solution is possibly a fix point of the Hamiltonian renormalization in \cite{Lang:2017beo}. Moreover this invariance indicates the scaling invariance from the viewpoint of lattice field theory, and relates to the conformal invariance in 3 dimensions.



\section{Improved Hamiltonian}\label{Improved Hamiltonian}

Our model is defined on a cubic lattice $\g$ which may be finite or infinite. For the purpose of relating LQC at $k=0$, we consider $\g$ to be a partition of 3-torus. The dynamical variables on $\g$ are holonomies and gauge covariant fluxes defined at all edges $e\in E(\g)$:
\be
h(e):=\mathcal{P} \exp \int_{e} A, \quad p^{a}(e):=-\frac{1}{2 \beta a^{2}} \operatorname{tr}\left[\tau^{a} \int_{S_{e}} \varepsilon_{i j k} d \sigma^{i} \wedge \mathrm{d} \sigma^{j} h\left(\rho_{e}(\sigma)\right) E_{b}^{k}(\sigma) \tau^{b} h\left(\rho_{e}(\sigma)\right)^{-1}\right]
\ee
where $S_e$ is a 2-face in the dual lattice $\g^*$, and $\rho_e(\sig)\subset S_e$ is a path starting at the begin point of $e$ and traveling along $e$ until $e\cap S_e$, then running in $S_e$ until $x$. $a$ is a length unit for making $p^a(e)$ dimensionless. $h(e),p^a(e)$ satisfies the holonomy-flux algebra
\be
\lt\{{h}(e),{h}(e')\rt\} &=&0\nonumber\\
\lt\{{p}^a(e),{h}(e')\rt\} &=&\frac{\kappa}{a^2} \delta_{e,e'} \frac{\t^a}{2} {h}(e')\nonumber\\
\lt\{{p}^a(e),{p}^b(e')\rt\}&=&-\frac{\kappa}{a^2} \delta_{e,e'} \eps_{abc} {p}^c(e'). \label{ph}
\ee
where $\kappa=16\pi G_N$. Our work is developed from the reduced phase space LQG where $h(e),p^a(e)$ are all physical Dirac observables.

\begin{figure}[t]
  \begin{center}
  \includegraphics[width = 0.5\textwidth]{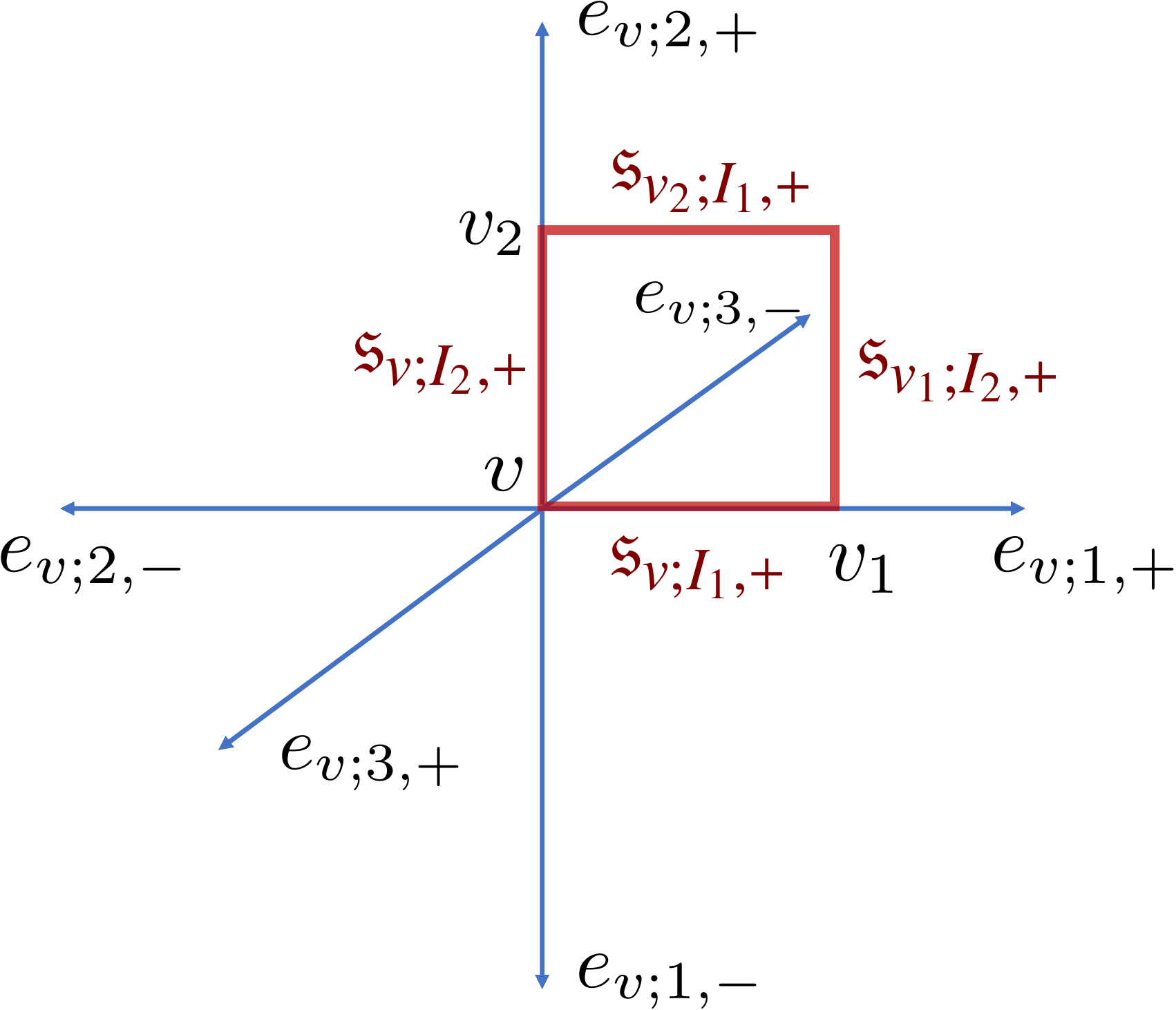}
  \end{center}
  \caption{A neighborhood of a vertex $v$ in the cubic lattice $\g$. The square bounded by red edges is $S(\Delta)$ of Planckian area $\Delta$. Loop holonomy $h(\Delta)$ is along $\partial S(\Delta)$.}
  \label{lattice}
  \end{figure}

We focus on the deparametrized model of gravity coupled to dust, the discrete physical Hamiltonian on $\g$ can be written as (see e.g. \cite{Han:2019vpw})
\be
{\bf H}&=&\sum_{v\in V(\g)}{\bf H}_v,\quad {\bf H}_v=\sqrt{C_{v}^2-\frac{\a}{4}C_{j,v}^2}\label{H}\\
C_{v}&=&-\frac{1}{\b^2}C_{0,v}-\frac{1+\b^2}{\b^2\kappa}\, {}^3 \calr_v+\frac{\L}{\kappa} V_v,
\ee
where $\a=1,0$ corresponds to the Brown-Kucha\v{r} or Gaussian dust. There are non-holonomic constraint $C_{v}<0$ and $C_{v}^2-\frac{\a}{4}C_{j,v}^2\geq0$. We focus on the physical dust with positive energy density, and the physical time flow is backward to make ${\bf H}$ positive (see Appendix \ref{BK} for details). We have include the cosmological constant term $\L V_v$ in the Hamiltonian. Here we employ the regularization of $C_v$ in \cite{Alesci:2014aza,Assanioussi:2015gka}. $^3 \mathcal{R}_{v}$ is the discrete 3-curvature:
\be
^3 \mathcal{R}_{v}&=&\sum_{I \neq J} \sum_{s_{1}, s_{2}=\pm 1} L_{v}\left(I, s_{1} ; J, s_{2}\right)\left(\frac{2 \pi}{\alpha}-\pi+\arccos \left[\frac{\vec{p}\left(e_{v ; I s_{1}}\right) \cdot \vec{p}\left(e_{v ; J s_{2}}\right)}{p\left(e_{v ; I s_{1}}\right) p\left(e_{v ; J s_{2}}\right)}\right]\right),\\
L_{v}\left(I, s_{1} ; J, s_{2}\right)&=&\frac{1}{V_{v}} \sqrt{\varepsilon^{a b c} p_{b}\left(e_{v ; I s_{1}}\right) p_{c}\left(e_{v ; J s_{2}}\right) \varepsilon^{a b^{\prime} c^{\prime}} p_{b^{\prime}}\left(e_{v ; I s_{1}}\right) p_{c^{\prime}}\left(e_{v ; J s_{2}}\right)}.
\ee 
In our notation, $e_{v;I,s}$ with $I=1,2,3$, $s=\pm$, and vertex $v\in V(\g)$ denotes an edge starting at $v$ oriented toward the ($I,s$) direction (see FIG.\ref{lattice}).

The Euclidean Hamiltonian and diffeomorphism constraints ${C}_{0,v}$, $C_{j,v}$ are written as 
\be
{C}_{\mu,v}&=&\frac{-4}{3\b\kappa^2}\sum_{s_1,s_2,s_3=\pm1}s_1s_2s_3\ \eps^{I_1I_2I_3}\, \mathrm{Tr}\Big[\t_\mu\ \Big({h}(\Box_{v;I_1s_1,I_2s_2})-{h}(\Box_{v;I_1s_1,I_2s_2})^{-1}\Big)\nonumber\\
&&\times\ {h}(e_{v;I_3s_3})\Big\{{h}(e_{v;I_3s_3})^{-1},{V}_v\Big\}\ \Big].\label{C}
\ee
where $\mu=0,1,2,3$ and $\t_\mu=({\bf 1},\t_j)$ and $\t_j=-i(\text{Pauli matrix})_j$. ${V}_v$ is the 3-volume at $v$:
\be
{V}_{v}&=&\left({Q}_{v}\right)^{1 / 2}, \\
 {Q}_{v}&=&\beta^{3} a^{6} \varepsilon^{a b c} \frac{{p}_{a}\left(e_{v ; 1+}\right)-{p}_{a}\left(e_{v ; 1-}\right)}{2} \frac{{p}_{b}\left(e_{v ; 2+}\right)-{p}_{b}\left(e_{v ; 2-}\right)}{2}\frac{{p}_{c}\left(e_{v ; 3+}\right)-{p}_{c}\left(e_{v ; 3-}\right)}{2}.
\ee

${\bf H}$ depends on 2 types of variables: flux $p^a(e)$ and loop holonomies $h(\Box)$, while $h(e)\{h(e)^{-1},V_v\}$ leads to the expression only involving fluxes:
\be
h(e_{v;J,s})\lt\{h(e_{v;J,s})^{-1},V^{}_v\rt\}
&=&\frac{s\,\b^3 a^4\kappa}{2Q_v^{1/2}}\eps^{JMN}\eps_{abc}\lt[\frac{\t^a/2}{2}\rt]\frac{X^b_M(v)}{2}\frac{X^c_N(v)}{2},\label{hhv}\\
X^b_M(v)&=&{p^b(e_{v;M,+})-p^b(e_{v;M,-})}.
\ee
Note that because $p^a(e)$ is covariant flux, we have
\be
p^{a}\left(e_{v ; I,-}\right)=\frac{1}{2} \operatorname{Tr}\left[\tau^{a} h\left(e_{v-\hat{I} ; I,+}\right)^{-1} p^{b}\left(e_{v-\hat{I} ; I,+}\right) \tau^{b} h\left(e_{v-\hat{I} ; I,+}\right)\right].
\ee




In the following, we are going to construct a new Hamiltonian by modifying $C_{\mu,v}$. The new Hamiltonian is resulting from a different regularization of the curvature $F_{ij}(A)$ in $C_{\mu,v}$. The new regularization replaces the loop holonomy $h(\Box)$ to a different expression: 

Firstly $h(\Box)$ relates to $F_{ij}$ when the scale of $\Box$ is small, 
\be
\frac{h(\Box)-h(\Box)^{-1}}{2}&\simeq& \half\int_\Box F_{ij}\, \rmd \sig^i\wedge\rmd\sig^j=\int_\Box F_{12}\, \rmd \sig^1\wedge\rmd\sig^2,\label{trick0}
\ee
where we use $\Box=\Box_{v;1,2}$ as an example.
We multiply and divide the integrand $\sqrt{\det(\fh)}$ where $\fh$ is the induced metric on $\Box$, and assume $\Box$ is small enough so that $F_{12}$ and ${\det (\fh)}$ are approximately constants. 
\be
\frac{h(\Box)-h(\Box)^{-1}}{2}&\simeq&\int_\Box \frac{F_{12}}{\sqrt{\det (\fh)}} \sqrt{\det (\fh)}\,\rmd \sig^1\wedge\rmd\sig^2\simeq \frac{F_{12}}{\sqrt{\det (\fh)}}(v)\int_\Box  \sqrt{\det (\fh)}\,\rmd \sig^1\wedge\rmd\sig^2\nonumber\\
&=&\frac{F_{12}}{\sqrt{\det (\fh)}}(v)\mathrm{Ar}(\Box)=\frac{F_{12}(v)\Delta}{\sqrt{\det (\fh(v))}}\frac{\mathrm{Ar}(\Box)}{\Delta},\label{trick1}
\ee
where $\fh$ is the induced metric on the 2d plaquette $\Box$. In the last step, we multiply and divide a quantum area scale $\Delta\simeq \ell_P^2=\hbar\kappa $ which may be chosen as the LQG minimal area gap: $\Delta=\frac{\sqrt{3}}{8}\b\ell_P^2$. We may write $\Delta=\int_{S(\Delta)}\sqrt{\det (\fh)}\,\rmd \sig^1\wedge\rmd\sig^2$ where $S(\Delta)$ is a small 2d surface of Planckian size (the square bounded by red edges in FIG.\ref{lattice}). We assume that sizes of $S(\Delta)$ and $\Box$ are similar, so $F_{12}$ and ${\det (\fh)}$ are also approximately constants on $S(\Delta)$. We can repeat the trick in Eq.\Ref{trick1},
\be
\frac{h(\Box)-h(\Box)^{-1}}{2}&\simeq&\frac{\mathrm{Ar}(\Box)}{\Delta}\frac{F_{12}(v)}{\sqrt{\det (\fh(v))}}\int_{S(\Delta)}\sqrt{\det (\fh)}\,\rmd \sig^1\wedge\rmd\sig^2\nonumber\\
&\simeq&\frac{\mathrm{Ar}(\Box)}{\Delta}\int_{S(\Delta)}F_{12}\,\rmd \sig^1\wedge\rmd\sig^2\simeq \frac{\mathrm{Ar}(\Box)}{\Delta}\frac{h(\Delta)-h(\Delta)^{-1}}{2}, \label{trick2}
\ee
where $h(\Delta)$ is the holonomy along the boundary of $S(\Delta)$. Here we set 
\be
{\rm Ar}(\Box_{v;I_1s_1,I_2s_2})=\half\b a^2\sum_{s_3=\pm}\sqrt{p^a(e_{v;I_3,s_3})p^a(e_{v;I_3,s_3})}\quad(I_3\neq I_1,I_2),
\ee 
and $\Ar(\Box)\sim\Delta\sim \ell_P^2$ so that ${\mathrm{Ar}(\Box)}/{\Delta}$ is finite. The error from approximations \Ref{trick0} - \Ref{trick2} is bounded by $O(\ell_P^3)$\footnote{Given any 2d integral of a bounded function $f(x)$ whose upper and lower bounds are $f(a)=f_{max},f(b)=f_{min}$, we have $|\int_S\rmd^2x\, f(x)-\int_S\rmd^2x\, f(v)|\leq \int_S\rmd^2x |f(x)- f(v)|\leq\Ar(S)|f_{max}-f_{min}|\leq C\Ar(S)||a-b||$ for some constant $C$, where $\Ar(S)\sim \ell_P^2$ and $||a-b||\sim\ell_P$ for $S=S(\Delta)$ or $\Box$. }, while $h(\Delta)-h(\Delta)^{-1}\sim O(\ell_P^2)$. As an interesting perspective, the dependence on lattice $\g$ is only reflected by the plaquette area ${\rm Ar}(\Box)$, while $h(\Delta)$ is independent of $\g$ thus is invariant under lattice refinement. The price is that the free parameter $\Delta$ has to be introduced. However it seems to us that $\Delta$ labels the ultraviolet (UV) energy scale where this theory is defined.

We define a new Hamiltonian by using \Ref{trick2},
\be
{\bf H}_\Delta&=&\sum_{v\in V(\g)}{\bf H}_{\Delta,v},\quad {\bf H}_{\Delta,v}=\sqrt{\lt(C_{v}^\Delta{}\rt)^2-\frac{\a}{4}\lt(C_{j,v}^\Delta\rt)^2}\label{HDelta}\\
C^\Delta_{v}&=&-\frac{1}{\b^2}C^\Delta_{0,v}-\frac{1+\b^2}{\b^2\kappa}\, {}^3 \calr_v+\frac{\L}{\kappa} V_v\\
{C}^\Delta_{\mu,v}&=&\frac{-4}{3\b\kappa^2}\sum_{s_1,s_2,s_3=\pm1}s_1s_2s_3\ \eps^{I_1I_2I_3}\frac{{\rm Ar}(\Box_{v;I_1s_1,I_2s_2})}{\Delta}\mathrm{Tr}\Big[\t_\mu\ \Big({h}(\Delta_{v;I_1s_1,I_2s_2})-{h}(\Delta_{v;I_1s_1,I_2s_2})^{-1}\Big)\ \nonumber\\
&&\times\ {h}(e_{v;I_3s_3})\Big\{{h}(e_{v;I_3s_3})^{-1},{V}_v\Big\}\ \Big]\label{CDelta}.
\ee
The derivation in Eq.\Ref{trick2} assumes that the size of $\Box$ is comparable to $S(\Delta)$. This indicates that ${\bf H}_\Delta$ is defined at UV or the deep quantum level where ${\rm Ar}(\Box)\sim \Delta\sim \ell_P^2$. In the semiclassical limit $\Delta\sim\ell_P^2\to0$ (or $\Delta\ll \mathrm{Ar}(\Box)$), the loop $\partial S(\Delta)$ shrinks and $h(\Delta)-h(\Delta)^{-1}\to 2\int_{S(\Delta)} F$. Then we reverse the derivation in Eqs.\Ref{trick1} and \Ref{trick2} and recover the classical expression of the physical Hamiltonian. If we keep ${\rm Ar}(\Box)\sim \Delta\sim \ell_P^2$ in the semiclassical limit (so $\Box$ also shrinks), we recover the continuum expression of the physical Hamiltonian. 


\section{Holonomy along Planckian Segment}\label{Holonomy along Planckian Segment}

As preparation for defining $h(\Delta)$ in ${\bf H}_\Delta$, in this section we construct the length-dependent holonomy $h_\Delta\lt(\fs\rt)$ along the segment $\fs$ whose length is fixed as $\sqrt{\Delta}$. We denote by $h_\Delta\lt(\fs_{v;I,s}\rt)$ with $s=+/-$ the holonomy along $\fs$ toward the positive/negative $I$-th direction at $v$. $\fs_{v;I,s}\subset e_{v;I,s}$ and shares the same source $v$ with $e_{v;I,s}$ ($\fs_{v;I,s}$ is the red segment along $e_{v;I,s}$ in FIG.\ref{lattice}). In contrast to $h\lt(e_{v;I,s}\rt)$, $h_\Delta\lt(\fs_{v;I,s}\rt)$ has $\fs_{v;I,s}$ with fixed length $\sqrt{\Delta}$:
\be
h_{\Delta}\lt(\fs_{v;I,s}\rt)=\mathcal{P}\exp\left[\int_{0}^{1}du\sum_{J=1}^3\frac{d\sig^{J}}{du}A_{J}^{a}(\vec{\sig}(u))\frac{\tau^{a}}{2}\right],\quad \sqrt{\Delta}=\int_{0}^{1}du\sqrt{q_{IJ}\frac{d\sig^{I}}{du}\frac{d\sig^{J}}{du}},\label{hDelta1}
\ee
where $\sig^J(u)$ ($J=1,2,3$) is a parametrization the segment $\fs_{v;I,s}$ such that the source and target of the segment correspond to $u=0,1$. For simplicity, we adapt the coordinates $\sig^J$ to the lattice $\g$, such that the $\sig^J$ coordinate axis is along $e_{v;I,s}$ or $\fs_{v;I,s}$. The tangent vector of $\fs_{v;I,s}$ has the only nonzero component $d \sig^I/du$ ($d \sig^J/d u=0$ for $J\neq I$). We assume $d \sig^I/du>0$ without loss of generality. Eq.\Ref{hDelta1} simplifies to
\be
h_{\Delta}(\fs_{v;I,s})&=&\mathcal{P}\exp\left[\int_{0}^{1}du\frac{d\sig^{I}}{du}A_{I}^{a}(\sig^{I}(u),0,0)\frac{\tau^{a}}{2}\right]\quad (\text{no sum in $I$}),\label{hDelta2}\\
\sqrt{\Delta}&=&\int_{0}^{1}du \frac{d\sig^{I}}{du}\sqrt{q_{II}}\label{sqrtDelta}
\ee
Eq.\Ref{sqrtDelta} can be solved by
\be
\frac{d\sig^{I}}{du}=\sqrt{\frac{\Delta}{q_{II}}}=\sqrt{\frac{\Delta\det(q)}{\frac{1}{2}\epsilon_{Imn}E_{b}^{m}E_{c}^{n}\epsilon_{Ipq}E_{b}^{p}E_{c}^{q}}}\quad (\text{no sum in $I$}),\label{solmubar}
\ee
where we have expressed the metric component $q_{II}=e_I^ae_I^a$ in terms of densitized triads $E_{a}^{j}$, using $e_{i}^{a}=\frac{1/2}{\sqrt{\det(q)}}\epsilon_{ijk}\epsilon^{abc}E_{b}^{j}E_{c}^{k}$. It is clear that the solution to Eq.\Ref{sqrtDelta} is not unique. We make the choice Eq.\Ref{solmubar} and insert in Eq.\Ref{hDelta2} as a definition of $h_{\Delta}(\fs)$. Moreover, we regularize ${d\sig^{I}}/{du}$ to express in terms of fluxes: For instance
\be
\frac{d\sig^{1}}{du}&=&\sqrt{\frac{\Delta\det(q)}{\left(E^{2}\cdot E^{2}\right)\left(E^{3}\cdot E^{3}\right)-\left(E^{2}\cdot E^{3}\right)^{2}}}\nonumber\\
&=&\delta\sig^1\sqrt{\frac{\Delta\det(q)}{\left(E^{2}\cdot E^{2}\right)\left(E^{3}\cdot E^{3}\right)-\left(E^{2}\cdot E^{3}\right)^{2}}}\frac{\delta\sig^{1}\delta\sig^{2}\delta\sig^{3}}{\delta\sig^{1}\delta\sig^{3}\delta\sig^{1}\delta\sig^{2}}\nonumber\\
&\simeq&\l\sqrt{\frac{ Q_v/\left(\beta^{3}a^{6}\right)}{\left[p(e_{v;2,s})\cdot p(e_{v;2,s})\right]\left[p(e_{v;3,s})\cdot p(e_{v;3,s})\right]-\left[p(e_{v;2,s})\cdot p(e_{v;3,s})\right]^{2}}}\equiv\bar{\mu}_{v;1,s},\\
 \lambda&=&\frac{\delta\sig^{1}}{\beta^{1/2}a}\sqrt{\Delta},
\ee
where $\delta\sig^I$ is the coordinate scale of the 2-surface $S_{e_{v;I,s}}$ defining $p(e_{v;I,s})$. We set $\delta\sig^I=1$ since the coordinate length of $e_{v;I,s}$ is set to be 1. Thus $\l$ is a constant. It is clear that $\bar{\mu}_{v;1,s}$ reduces to ${d\sig^{1}}/{du}$ when taking the continuum limit. Similarly, we have
\be
\frac{d\sig^{2}}{du}\simeq \bar{\mu}_{v;2,s}=\l\sqrt{\frac{Q_v/\left(\beta^{3}a^{6}\right)}{\left[p(e_{v;1,s})\cdot p(e_{v;1,s})\right]\left[p(e_{v;3,s})\cdot p(e_{v;3,s})\right]-\left[p(e_{v;1,s})\cdot p(e_{v;3,s})\right]^{2}}},\\
\frac{d\sig^{3}}{du}\simeq \bar{\mu}_{v;3,s}=\l\sqrt{\frac{Q_v/\left(\beta^{3}a^{6}\right)}{\left[p(e_{v;1,s})\cdot p(e_{v;1,s})\right]\left[p(e_{v;2,s})\cdot p(e_{v;2,s})\right]-\left[p(e_{v;1,s})\cdot p(e_{v;2,s})\right]^{2}}}.
\ee

We obtain the following result when inserting the above $\bar{\mu}_{v;I,s}$ in $h_\Delta(\fs_{v;I,s})$:

\begin{Lemma}\label{lemmamubar}

A gauge transformation can lead to the following expression for $h_\Delta(\fs_{v;I,s})$ while leaving $h(e_{v;I,s})$ invariant:
\be
h_\Delta(\fs_{v;I,s})=e^{\bar{\mu}_{v;I,s}\theta^a(e_{v;I,s})\t^a/2},\quad \fs_{v;I,s}\subset e_{v;I,s}\label{hstheta}
\ee
where $\theta^a(e_{v;I,s})$ is given by the lattice edge holonomy $h(e_{v;I,s})$:
\be
h(e_{v;I,s})=e^{\theta^a(e_{v;I,s})\t^a/2}.
\ee

\end{Lemma}

{\bf Proof:} We impose locally the following axial gauge to the connection $A$ along every edge $e_{v;I,s}$, i.e. at a given $e_{v;I,s}$, the restriction of $A$ along the edge, $A_I^a$, is fixed to be a constant.  
\be
A_I^a(\vec{\sig})=\theta^a(e_{v;I,s}), \quad \vec{\sig}\in e_{v;I,s},\label{axial}
\ee
Indeed Eq.\Ref{axial} can be obtained by a gauge transformation from generic $A$: 
\be
g({\sig}^I)A_I({\sig}^I)g({\sig}^I)^{-1}-\partial_I g({\sig}^I)\,g({\sig}^I)^{-1}=\theta(e_{v;I,s}),
\ee
where $A_I({\sig}^I)=A^a_I({\sig}^I)\frac{\t^a}{2}$ and $ \theta(e_{v;I,s})=\theta^a(e_{v;I,s})\frac{\t^a}{2}$. Equivalently, the above gauge transformation is a 1st order ordinary differential equation of $g(\sig^I)$ along $e_{v;I,s}$:
\be
\partial_I g({\sig}^I)=g(\sig^I)A_I({\sig}^I)-\theta(e_{v;I,s})g(\sig^I).
\ee
The solution to the above equation is given by\footnote{The holonomy $h(\sig^I)\equiv\mathcal{P}\exp\left[\int_{0}^{\sig^I}d\sig'^{I} A_{I}(\sig'^I)\right]$ satisfies $\partial_I h(\sig^I)=h(\sig^I)A_{I}(\sig^I)$.}
\be
g({\sig}^I)=e^{-\sig^I\theta(e_{v;I,s})}\,\mathcal{P}\exp\left[\int_{0}^{\sig^I}d\sig'^{I} A_{I}(\sig'^I)\right].\label{gaugetrans}
\ee
If the constant $\theta(e_{v;I,s})$ is chosen such that
\be
e^{\theta(e_{v;I,s})}=\mathcal{P}\exp\left[\int_{0}^{1}d\sig'^{I} A_{I}(\sig'^I)\right]=h(e_{v;I,s}),
\ee
we obtain that gauge transformations are identities at the source and target of $e_{v;I,s}$.
\be
g(0)=g(1)=1.
\ee
Therefore the gauge transformation $g(\vec{\sig})$ leaves all lattice edge holonomies $h(e_{v;I,s})$ invariant, while transforming locally the connection $A$. Inserting Eq.\Ref{axial} and the definition of $\bar{\mu}_{v;I,s}$ into $h_\Delta(\fs_{v;I,s})$ gives Eq.\Ref{hstheta}.\\
$\Box$

\section{Loop Holonomy around Planckian Plaquette}\label{Loop Holonomy around Planckian Plaquette}

The curvature $F(A)$ smeared on the Planckian size $S(\Delta)$ relates to the loop holonomy $h(\Delta)$ along $\partial S(\Delta)$. Here we assume the shape of $S(\Delta)$ to be a square, i.e. $\partial S(\Delta)$ is made by 4 edges (see FIG.\ref{lattice}), and $h(\Delta)$ is given by
\be
h\lt(\Delta_{v;I_1,s_1,I_2,s_2}\rt)=h_\Delta\lt(\fs_{v;I_1,s_1}\rt){h}_\Delta\lt(\fs_{v_1;I_2,s_2}\rt){h}_\Delta\lt(\fs_{v_2;I_1,s_1}\rt)^{-1}h_\Delta\lt(\fs_{v;I_2,s_2}\rt)^{-1}.
\ee
where $v_1,v_2$ are vertices of $\partial S(\Delta)$ (see FIG.\ref{lattice}). $h_\Delta\lt(\fs_{v;I_1,s_1}\rt)$ and $h_\Delta\lt(\fs_{v;I_2,s_2}\rt)$ with $\fs_{v;I_1,s_1}\subset e_{v;I_1,s_1}$ and $\fs_{v;I_2,s_2}\subset e_{v;I_2,s_2}$ are given by Eq.\Ref{hstheta}. However ${h}_\Delta\lt(\fs_{v_1;I_2,s_2}\rt)$ and ${h}_\Delta\lt(\fs_{v_2;I_1,s_1}\rt)$ are based on segments $\fs_{v_1;I_2,s_2},\fs_{v_2;I_1,s_1}$ which do not belong to any edge in $\g$. To construct these 2 holonomies, we write
\be
{h}_\Delta\lt(\fs_{v_2;I_1,s_1}\rt)=e^{X_1},\quad {h}_\Delta\lt(\fs_{v_1;I_2,s_2}\rt)=e^{X_2},
\ee
and expand $h\lt(\Delta\rt)$ since the size of $S(\Delta)$ is Planckian: 
\be
h\lt(\Delta_{v;I_1,s_1,I_2,s_2}\rt)&=&1+\lt(\bar{\mu}_{v;I_1,s_1}\theta(e_{v;I_1,s_1})-X_1\rt)-\lt(\bar{\mu}_{v,I_2,s_2}\theta(e_{v;I_2,s_2})-X_2\rt)\nonumber\\
&&+\bar{\mu}_{v;I_1,s_1}\bar{\mu}_{v,I_2,s_2}\lt[\theta(e_{v;I_1,s_1}),\,\theta(e_{v;I_2,s_2})\rt]+O(\ell_P^3)
\ee
where we have ignored higher orders in size of $S(\Delta)$. $h\lt(\Delta_{v;I_1,s_1,I_2,s_2}\rt)-1$ approximates the curvature integrated on $S(\Delta)$
\be
&&\lt(\bar{\mu}_{v;I_1,s_1}\theta(e_{v;I_1,s_1})-X_1\rt)-\lt(\bar{\mu}_{v,I_2,s_2}\theta(e_{v;I_1,s_1})-X_2\rt)+\bar{\mu}_{v;I_1,s_1}\bar{\mu}_{v,I_2,s_2}\lt[\theta(e_{v;I_1,s_1}),\,\theta(e_{v;I_2,s_2})\rt]\nonumber\\
&\simeq& \frac{1}{2} \int_{S(\Delta_{v;I_1,s_1,I_2,s_2})}\rmd\sig^I\wedge \rmd\sig^J\,F_{IJ}(\vec{\sig}).
\label{mumuF}
\ee

On the other hand, the loop holonomy along the lattice edges can be expanded similarly:
\be
h\lt(\Box_{v;I_1,s_1,I_2,s_2}\rt)&=&h\lt(e_{v;I_1,s_1}\rt){h}\lt(e_{v+s_1\hat{I}_1;I_2,s_2}\rt){h}\lt(e_{v+s_2\hat{I}_2;I_1,s_1}\rt)^{-1}h\lt(e_{v;I_2,s_2}\rt)^{-1}\nonumber\\
&\simeq &1+\lt(\theta(e_{v;I_1,s_1})-\theta(e_{v+s_2\hat{I}_2;I_1,s_1})\rt)-\lt(\theta(e_{v;I_2,s_2})-\theta(e_{v+s_1\hat{I}_1;I_2,s_2})\rt)\nonumber\\
&&+\lt[\theta(e_{v;I_1,s_1}),\,\theta(e_{v;I_2,s_2})\rt]
\ee
up to higher order in lattice size. The curvature integrated on the lattice plaquette $\Box$ is approximated by $h\lt(\Box_{v;I_1,s_1,I_2,s_2}\rt)-1$:
\be
\frac{1}{2} \int_{\Box_{v;I_1,s_1,I_2,s_2}}\rmd\sig^I\wedge \rmd\sig^J\,F_{IJ}(\vec{\sig})&\simeq&\lt(\theta(e_{v;I_1,s_1})-\theta(e_{v+s_2\hat{I}_2;I_1,s_1})\rt)-\lt(\theta(e_{v;I_2,s_2})-\theta(e_{v+s_1\hat{I}_1;I_2,s_2})\rt)\nonumber\\
&&+\lt[\theta(e_{v;I_1,s_1}),\,\theta(e_{v;I_2,s_2})\rt].\label{F}
\ee
Consistent with our regularization \Ref{trick1} and \Ref{trick2}, $F_{IJ}(\vec{\sig})$ are approximately constant on both $S(\Delta)$ and $\Box$. Therefore
\be
\int_{S(\Delta_{v;I_1,s_1,I_2,s_2})}\rmd\sig^I\wedge \rmd\sig^J\,F_{IJ}(\vec{\sig})\simeq\bar{\mu}_{v;I_1,s_1}\bar{\mu}_{v,I_2,s_2}\int_{\Box_{v;I_1,s_1,I_2,s_2}}\rmd\sig^I\wedge \rmd\sig^J\,F_{IJ}(\vec{\sig}).
\ee
This relation and comparing Eqs.\Ref{mumuF} and \Ref{F} suggest the following definition of $X_1$ and $X_2$
\be
X_1:=\bar{\mu}_{v;I_1,s_1}\theta(e_{v;I_1,s_1})-\bar{\mu}_{v;I_1,s_1}\bar{\mu}_{v;I_2,s_2}\lt(\theta(e_{v;I_1,s_1})-\theta(e_{v+s_2\hat{I}_2;I_1,s_1})\rt),\\
X_2:=\bar{\mu}_{v;I_2,s_2}\theta(e_{v;I_2,s_2})-\bar{\mu}_{v;I_1,s_1}\bar{\mu}_{v;I_2,s_2}\lt(\theta(e_{v;I_2,s_2})-\theta(e_{v+s_1\hat{I}_1;I_2,s_2})\rt).
\ee
With the above $X_1, X_2$, the holonomies ${h}_\Delta\lt(\fs_{v_2;I_1,s_1}\rt)=e^{X_1}$, ${h}_\Delta\lt(\fs_{v_1;I_2,s_2}\rt)=e^{X_2}$
reduce to ${h}_\Delta\lt(\fs_{v;I_1,s_1}\rt)$, $ {h}_\Delta\lt(\fs_{v;I_2,s_2}\rt)$ when the connection $A$ is constant on $\Box$, and reduce to ${h}(e_{v+s_1\hat{I}_1;I_2,s_2})$, ${h}(e_{v+s_2\hat{I}_2;I_1,s_1})$ when $\bar{\mu}_{v;I_1,s_1}=\bar{\mu}_{v;I_2,s_2}=1$, i.e. when $\Box$ coincides with $S(\Delta)$.

To interprete ${h}_\Delta\lt(\fs_{v_1;I_2,s_2}\rt)$ and ${h}_\Delta\lt(\fs_{v_2;I_1,s_1}\rt)$ as holonomies, we define the following connection field in the plaquette $\Box_{v,I,J}$ bounded by $e_{v;I_1,s_1}$ $e_{v;I_2,s_2}$ ($I,J=1,2,3$)
\be
A(\vec{\sig})&=&\lt[(1-\sig^{I_2})\theta(e_{v;I_1,s_1})+\sig^{I_2} \theta(e_{v+s_2\hat{I}_2;I_1,s_1})\rt]\rmd\sig^{I_1}\nonumber\\
&&+\ \lt[(1-\sig^{I_1})\theta(e_{v;I_2,s_2})+\sig^{I_1} \theta(e_{v+s_1\hat{I}_1;I_2,s_2})\rt]\rmd\sig^{I_2}.\label{connA}
\ee
$A_{I_1}$ reduces to $\theta(e_{v;I_1,s_1})$ (or $\theta(e_{v+s_2\hat{I}_2;I_1,s_1})$) along $e_{v;I_1,s_1}$ at $\sig^{I_2}=0$ (or $e_{v+s_2\hat{I}_2;I_1,s_1}$ at $\sig^{I_2}=1$), while $A_{I_2}$ reduces to $\theta(e_{v;I_2,s_2})$ (or $\theta(e_{v+s_1\hat{I}_1;I_2,s_2})$) along $e_{v;I_2,s_2}$ at $\sig^{I_1}=0$ (or $e_{v+s_1\hat{I}_1;I_2,s_2}$ at $\sig^{I_1}=1$). Therefore $A$ is an extension of the connection along lattice edges to the plaquette $\Box$. It may be generalized to a connection in the 3d cube by 
\be
A(\vec{\sig})&=&\lt[(1-\sig^{2}-\sig^3)\theta(e_{v;1,+})+\sig^{2} \theta(e_{v+\hat{2};1,+})+\sig^{3} \theta(e_{v+\hat{3};1,+})\rt]\rmd\sig^{1}\nonumber\\
&&+\ \lt[(1-\sig^{1}-\sig^3)\theta(e_{v;2,+})+\sig^1 \theta(e_{v+\hat{1};2,+})+\sig^3 \theta(e_{v+\hat{3};2,+})\rt]\rmd\sig^{2}\nonumber\\
&&+\ \lt[(1-\sig^{1}-\sig^2)\theta(e_{v;3,+})+\sig^1 \theta(e_{v+\hat{1};3,+})+\sig^2 \theta(e_{v+\hat{2};3,+})\rt]\rmd\sig^{3}.\label{connA3}
\ee.

The above proposed relation between $A$ and $\theta(e)$ is not gauge invariant. One may notice that $A$ in Eq.\Ref{connA3} satisfy $\sum_{I=1}^3\partial_IA_I=0$ which cannot be always satisfied by all gauge transformations of $A$. Therefore the expression of $A$ in Eq.\Ref{connA} or \Ref{connA3} in terms of $\theta(e)$ depends on certain gauge fixing. Since we are going to apply ${\bf H}_{\Delta}$ to cosmology, we impose the following gauge fixing condition to $\theta(e)$
\be
\sum_{a=1}^3\delta_a^{I+1}\theta^{a}(e_{v;I,+})=0.\label{gauge condition}
\ee
In Section \ref{Homogeneous Effective Dynamics}, when we perturb $\theta^a(e_{v;I,+})$ from the homogeneous variables: $\theta^a(e_{v;I,+})=\lt[\theta_I+\delta \theta_\parallel(e_{v;I,+})\rt]\delta^a_I+\delta \theta_\bot^a(e_{v;I,+})$ where $\delta \theta_\bot^a(e_{v;I,+})$ is perpendicular to $\delta^a_I$, the gauge condition fixes one component of $\delta \theta_\bot^a(e_{v;I,+})$ to zero.

In terms of $\vec{\sig}$-coordinate, $v=(0,0)$, $v_1=(\bar{\mu}_{v,1,s_1},0)$, $v_2=(0, \bar{\mu}_{v,2,s_2})$, and $v^\star=(\bar{\mu}_{v,1,s_1}, \bar{\mu}_{v,2,s_2})$ where $v^\star$ is the intersection between $\fs_{v_1;I_2,s_2}$ and $\fs_{v_2;I_1,s_1}$. Along $\fs_{v_1;I_2,s_2}$ (or $\fs_{v_2;I_1,s_1}$), $A_{I_2}$ (or $A_{I_2}$) is constant:
\be
A_{I_2}(\bar{\mu}_{v;I_1,s_1},\sig^{I_2})&=&\lt[(1-\bar{\mu}_{v;I_1,s_1})\theta(e_{v;I_2,s_2})+\bar{\mu}_{v;I_1,s_1} \theta(e_{v+s_1\hat{I}_1;I_2,s_2})\rt] \quad \text{along}\ \fs_{v_1;I_2,s_2}\\
A_{I_1}(\sig^{I_1},\bar{\mu}_{v;I_2,s_2})&=&\lt[(1-\bar{\mu}_{v;I_2,s_2})\theta(e_{v;I_1,s_1})+\bar{\mu}_{v;I_2,s_2} \theta(e_{v+s_2\hat{I}_2;I_1,s_1})\rt] \quad \text{along}\ \fs_{v_2;I_1,s_1}
\ee
Therefore the holonomies along $\fs_{v_1;I_2,s_2}$ and $\fs_{v_2;I_1,s_1}$ reproduces
\be
{h}_\Delta\lt(\fs_{v_1;I_2,s_2}\rt)=\mathcal{P}\exp\left[\int_{0}^{1}du\frac{d\sig^{I_2}}{du}A_{I_2}\lt(\bar{\mu}_{v;I_1,s_1},\sig^{I_2}\rt)\right]=e^{X_2},\\
{h}_\Delta\lt(\fs_{v_2;I_1,s_1}\rt)=\mathcal{P}\exp\left[\int_{0}^{1}du\frac{d\sig^{I_1}}{du}A_{I_1}\lt(\sig^{I_1},\bar{\mu}_{v;I_2,s_2}\rt)\right]=e^{X_1}.
\ee
where $\frac{d\sig^{I}}{du}$ is the same as $\bar{\mu}_{v;I,s}$ at $v$. Similarly the connection also reproduce holonomies ${h}_\Delta\lt(\fs_{v;I_2,s_2}\rt)$ and ${h}_\Delta\lt(\fs_{v;I_1,s_1}\rt)$ in Eq.\Ref{hstheta}.

In summary, we have obtained the following definition of the loop holonomy $h(\Delta)$ around a Planckian size plaquette $S(\Delta)$:
\be
h\lt(\Delta_{v;I_1,s_1,I_2,s_2}\rt)=h_\Delta\lt(\fs_{v;I_1,s_1}\rt){h}_\Delta\lt(\fs_{v_1;I_2,s_2}\rt){h}_\Delta\lt(\fs_{v_2;I_1,s_1}\rt)^{-1}h_\Delta\lt(\fs_{v;I_2,s_2}\rt)^{-1}.
\ee
where holonomies along segments are given by
\be
h_\Delta\lt(\fs_{v;I_1,s_1}\rt)&=&e^{\bar{\mu}_{v;I_1,s_1}\theta(e_{v;I_1,s_1})},\label{hDelta1}\\
{h}_\Delta\lt(\fs_{v_1;I_2,s_2}\rt)&=&e^{\bar{\mu}_{v;I_2,s_2}\theta(e_{v;I_2,s_2})-\bar{\mu}_{v;I_1,s_1}\bar{\mu}_{v;I_2,s_2}\lt(\theta(e_{v;I_2,s_2})-\theta(e_{v+s_1\hat{I}_1;I_2,s_2})\rt)},\\
{h}_\Delta\lt(\fs_{v_2;I_1,s_1}\rt)^{-1}&=&e^{-\bar{\mu}_{v;I_1,s_1}\theta(e_{v;I_1,s_1})+\bar{\mu}_{v;I_1,s_1}\bar{\mu}_{v;I_2,s_2}\lt(\theta(e_{v;I_1,s_1})-\theta(e_{v+s_2\hat{I}_2;I_1,s_1})\rt)},\\
h_\Delta\lt(\fs_{v;I_2,s_2}\rt)^{-1}&=&e^{-\bar{\mu}_{v;I_2,s_2}\theta(e_{v;I_2,s_2})}\label{hDelta4}.
\ee
$h(\Delta)$ is the loop holonomy of the connection $A$ in Eq.\Ref{connA}. Although $h(\Delta)$ contains holonomies that do not along edges in the lattice $\g$, we are able to expresses $h(\Delta)$ in terms of lattice variables $\theta(e)=\theta^a(e)\t^a/2$ and $p^a(e)$ by the above construction. Inserting the above loop holonomy into Eqs.\Ref{HDelta} and \Ref{CDelta} defines the improved Hamiltonian ${\bf H}_\Delta$.

Simple expressions of $h_\Delta(\fs)$ in terms of $\theta(e)$ (in Eqs.\Ref{hDelta1} - \Ref{hDelta4}) rely on the expression of $A$ (in Eq.\Ref{connA}) which depends on the gauge fixing. Although a generic gauge transformation leave ${\bf H}_\Delta$ invariant, it may change expressions of $h_\Delta(\fs)$ by adding terms with higher order in $\bar{\mu}$ to their exponents, as suggested by some numerical tests. 

Although ${\bf H}_\Delta$ with $h_\Delta(\fs)$ in Eqs.\Ref{hDelta1} - \Ref{hDelta4} depends on the gauge fixing condition, the effective dynamics of cosmology derived from ${\bf H}_\Delta$ turns out to be gauge invariant (Gauss constraint is preserved by the dynamics) and independent of gauge fixing.


\section{Effective Equations of Full LQG}\label{Path Integral and Effective Equations}

We propose a discrete (canonical) effective action which govern the effective dynamics of the full LQG
\be
S_{eff}[g, h]&=&\sum_{i=0}^{N+1} K\left(g_{i+1}, g_{i}\right)-\frac{i \kappa}{a^{2}} \sum_{i=1}^{N} \delta \tau{\bf H}_\Delta[g_i], 
\ee
where $g_i=\{g_i(e)\}_{e\in E(\g)}$, $g_i(e)=e^{-ip^a(e)\t^a/2}e^{\theta^a(e)\t^a/2}$, and $i=1,\cdots,N$ label steps of discrete time evolution $\delta\t$. $a$ is a length unit. The kinetic term in the action is implied by the coherent state path integral in \cite{Han:2019vpw}: 
\be
K\left(g_{i+1}, g_{i}\right)&=&\sum_{e \in E(\gamma)}\left[z_{i+1, i}(e)^{2}-\frac{1}{2} p_{i+1}(e)^{2}-\frac{1}{2} p_{i}(e)^{2}\right],\\
z_{i+1, i}(e)&=&\operatorname{arccosh}\left(x_{i+1, i}(e)\right), \quad x_{i+1, i}(e)=\frac{1}{2} \operatorname{tr}\left[g_{i+1}(e)^{\dagger} g_{i}(e)\right].
\ee
The effective action $S_{eff}$ is designed in analogy with the ``classical action'' in the path integral formula derived in \cite{Han:2019vpw}, while here we employ the improved Hamiltoninan ${\bf H}_\Delta$. Due to the dependence of the scale $\Delta\sim \ell^2_P$, $S_{eff}$ is viewed as an ``quantum effective action'' defined at the high energy scale corresponding to $\Delta$. $S_{eff}$ contains quantum effects and depends on the scale $\Delta$ as an analog of the UV cut-off in quantum field theory. Variables $p^a(e),\theta^a(e)$ are vacuum expectation values (VEVs) satisfying $\delta S_{eff}=0$. We expect that $S_{eff}$ might be derived from the path integral in \cite{Han:2019vpw} by the standard procedure of quantum effective action in quantum field theory.



The variational principle $\delta S_{eff}=0$ gives the following equation of motion to determine VEVs $p^a(e),\theta^a(e)$ \cite{Han:2019vpw}:
\begin{itemize}

\item For $i=1,\cdots,N$, at every edge $e\in E(\g)$,
\be
\frac{1}{\delta\t}\lt[\frac{z_{i+1,i}(e)\,\tr\lt[\t^a g_{i+1}(e)^\dagger g_i(e)\rt]}{\sqrt{x_{i+1,i}(e)-1}\sqrt{x_{i+1,i}(e)+1}}-\frac{p_i(e)\,\tr\lt[\t^a g_{i}(e)^\dagger g_i(e)\rt]}{\sinh(p_i(e))}\rt]
=\frac{i\kappa}{a^2}\frac{\partial\,{\mathbf{H}_\Delta\lt[g_i^\eps\rt]}}{\partial{\eps^a(e)}}\Bigg|_{\vec{\eps}=0}\label{eoms1}
\ee

\item For $i=2,\cdots,N+1$, at every edge $e\in E(\g)$,
\be
\frac{1}{\delta\t}\lt[\frac{z_{i,i-1}(e)\,\tr\lt[\t^a g_{i}(e)^\dagger g_{i-1}(e)\rt]}{\sqrt{x_{i,i-1}(e)-1}\sqrt{x_{i,i-1}(e)+1}}-\frac{p_i(e)\,\tr\lt[\t^a g_{i}(e)^\dagger g_i(e)\rt]}{\sinh(p_i(e))}\rt]
=-\frac{i\kappa}{a^2}\frac{\partial\,{\mathbf{H}_\Delta\lt[g_i^\eps\rt]}}{\partial{\bar{\eps}^a(e)}}\Bigg|_{\vec{\eps}=0}.\label{eoms2}
\ee


\end{itemize}
On the right-hand sides of Eqs.\Ref{eoms1} and \Ref{eoms2}, 
\be
g_{i}^{\varepsilon}(e)=g_{i}(e) e^{\varepsilon_{i}^{a}(e) \tau^{a}}.\label{perturb}
\ee

The above equations of motion are understood as quantum effective equations since they are derived from the quantum effective action $S_{eff}$. These effective equations determine the improved effective dynamics of the full LQG. These equations are complemented by the gauge condition for defining ${\bf H}_\Delta$.

\section{Homogeneous Effective Dynamics}\label{Homogeneous Effective Dynamics}

In order to make contact to LQC, we study solutions of effective equations which are homogeneous (non-isotropic) at every time-slice $\cs$. We impose the following ansatz at every $\cs$:
\be
p^a(e_{v;I,s})=s\, p_I(\t)\delta^a_I,\ (p_I>0), \quad \theta^a(e_{v;I,s})=s\, \theta_I(\t)\delta^a_I,\label{homo}
\ee
where $p_I(\t),\theta_I(\t)$ ($I=1,2,3$) are $6$ constants on $\cs$ but evolve in time. It is easy to see that four $h_\Delta(\fs)$'s in $h(\Delta)$ reduces to the holonomies in LQC $\bar{\mu}$-scheme:
\be
h_\Delta(\fs_{v;I,s})=\exp\lt(s \l \sqrt{\frac{p_I}{p_{I+1}p_{I+2}}}\theta_I\t_I/2\rt),
\ee
where $I,I+1,I+2$ are defined mod 3.

We insert the homogeneous ansatz \Ref{homo} into effective equations \Ref{eoms1} and \Ref{eoms2}, and take the continuous limit $\delta\t\to0$. The effective equations reduce to differential equations of homogeneous variables $p_I(\t),\theta_I(\t)$. A part of this computation is a simple generalization from the computation for homogeneous and isotropic cosmology in \cite{Han:2019vpw}, and is sketched as follows:

First of all, inserting \Ref{homo} into Eqs.\Ref{eoms1} and \Ref{eoms2} gives
\be
\delta_{I}^{a}\left[\frac{d\theta_I}{d\tau}+i \frac{d p_{I}}{d \tau}\right]=\left.\frac{i \kappa}{a^{2}} \frac{\partial \mathbf{H}_\Delta\left[g^{\varepsilon}\right]}{\partial \varepsilon^{a}\left(e_{I}(v)\right)}\right|_{\vec{\varepsilon}=0},\quad 
\delta_{I}^{a}\left[\frac{d\theta_I}{d\tau}-i \frac{d p_{I}}{d \tau}\right]=-\left.\frac{i \kappa}{a^{2}} \frac{\partial \mathbf{H}_\Delta\left[g^{\varepsilon}\right]}{\partial \bar{\varepsilon}^{a}\left(e_{I}(v)\right)}\right|_{\vec{\varepsilon}=0},\label{homoeom0}
\ee
where $e_I(v)\equiv e_{v;I,+}$. 

Since ${\bf H}_\Delta$ is conveniently expressed as a function of $p^a(e),\theta^a(e)$, we would like to write the right-hand sides of Eq.\Ref{homoeom0} in terms of derivatives of $p^a(e),\theta^a(e)$. Eq.\Ref{perturb} can be rewritten in the polar-decomposition form where we extract perturbations of $p^a,\theta^a$:
\be
&&g^\eps(e_I(v))=e^{\lt(\theta-{i}p\rt)\frac{\t^I}{2}}e^{\eps^a(e_I(v))\t^a}=e^{-ip^a(e_I(v))\t^a/2}e^{\theta^a(e_I(v))\t^a/2},\label{perturbeqn}
\ee
where $p^a,\theta^a$ contains longitudinal perturbations $\delta p_\parallel,\delta \theta_\parallel$ and transverse perturbations $\delta p_\bot^a,\delta \theta_\bot^a$ with $a=I+1,I+2$ mod $3$ ,
 \be
 p^a(e_I(v))&=&\lt[p_I+\delta p_\parallel(e_I(v))\rt]\delta^a_I+\delta p_\bot^a(e_I(v)),\label{deltap}\\
 \theta^a(e_I(v))&=&\lt[\theta_I+\delta \theta_\parallel(e_I(v))\rt]\delta^a_I+\delta \theta_\bot^a(e_I(v)).\label{deltatheta}
\ee
$\delta p_\parallel\delta^a_I,\delta \theta_\parallel\delta^a_I$ and $\delta p_\bot^a,\delta \theta_\bot^a$ are perpendicular and relate to $\eps^a$ up to $O(\eps^2)$ by 
\be
\delta p_\parallel(e_I(v))&=&i\lt[{\eps^I(e_I(v))-\bar{\eps}^I(e_I(v))}\rt],\quad 
\delta \theta_\parallel(e_I(v))={\eps^I(e_I(v))+\bar{\eps}^I(e_I(v))},\\
\left(\begin{array}{c}\delta p^{I+1}_\bot(e_I(v))\\ \delta p^{I+2}_\bot(e_I(v))\end{array}\right)&=&\frac{i p}{\sinh(p)}\left(\begin{array}{cc}\cos(\theta) & -\sin(\theta)  \\ \sin(\theta) & \cos(\theta) \end{array}\right) \left(\begin{array}{c}{\eps^{I+1}(e_I(v))-\bar{\eps}^{I+1}(e_I(v))}\\ {\eps^{I+2}(e_I(v))-\bar{\eps}^{I+2}(e_I(v))}\end{array}\right),\\
\left(\begin{array}{c}\delta \theta^{I+1}_\bot(e_I(v))\\ \delta \theta^{I+2}_\bot(e_I(v))\end{array}\right)&=&\frac{\theta/2}{\sin(\theta/2)}\Bigg[\left(\begin{array}{cc}\cos(\theta/2) & -\sin(\theta/2)  \\ \sin(\theta/2) & \cos(\theta/2) \end{array}\right) \left(\begin{array}{c}{\eps^{I+1}(e_I(v))+\bar{\eps}^{I+1}(e_I(v))}\\ {\eps^{I+2}(e_I(v))+\bar{\eps}^{I+2}(e_I(v))}\end{array}\right)\nonumber\\
&&+i\tanh(p/2)\left(\begin{array}{cc}\sin(\theta/2) & \cos(\theta/2)  \\ -\cos(\theta/2) & \sin(\theta/2) \end{array}\right)\left(\begin{array}{c}{\eps^{I+1}(e_I(v))-\bar{\eps}^{I+1}(e_I(v))}\\ {\eps^{I+2}(e_I(v))-\bar{\eps}^{I+2}(e_I(v))}\end{array}\right)\Bigg].\label{solvedeltaptheta}
\ee
The above linear transformation between $\eps^a$ and $\delta p,\delta\theta$ is non-degenerate. By changing variables, Eq.\Ref{homoeom0} reduces to
\begin{itemize}

\item The diagonals $a=I$ give time evolution equations
\be
\frac{\mathrm{d} \theta_I}{\mathrm{d} \tau}=-\left.\frac{\kappa}{a^{2}} \frac{\partial \mathbf{H}_\Delta\left[g^{\varepsilon}\right]}{\partial \delta p_{\|}\left(e_{I}(v)\right)}\right|_{\delta \theta=\delta p=0}, \quad \frac{\mathrm{d} p_I}{\mathrm{d} \tau}=\left.\frac{\kappa}{a^{2}} \frac{\partial \mathbf{H}_\Delta\left[g^{\varepsilon}\right]}{\partial \delta \theta_{\|}\left(e_{I}(v)\right)}\right|_{\delta \theta=\delta p=0}.\label{homeom1}
\ee

\item The off-diagonals $a\neq I$ give constraint equations
\be
\left.\frac{\partial \mathbf{H}_\Delta\left[g^{\varepsilon}\right]}{\partial \delta p_{\perp}^{a}\left(e_{I}(v)\right)}\right|_{\delta \theta=\delta p=0}=0,\left.\quad \frac{\partial \mathbf{H}_\Delta\left[g^{\varepsilon}\right]}{\partial \delta \theta_{\perp}^{a}\left(e_{I}(v)\right)}\right|_{\delta \theta=\delta p=0}=0, \quad a \neq I.\label{homeom2}
\ee

\end{itemize}

The gauge condition Eq.\Ref{gauge condition} set some components of $\delta\theta_\bot$ to zero. So Eq.\Ref{homeom2} only needs to take into account derivatives of nonzero $\delta\theta_\bot$. However we can show (in below) that Eq.\Ref{homeom2} is satisfied for all $\delta\theta_\bot$ even without imposing the gauge condition. Moreover the (discrete) Gauss constraint (namely, the closure condition)
\be
-\sum_{e, s(e)=v}p_1^a(e)+\sum_{e, t(e)=v}\L^a_{\ b}\lt(\vec{\theta}_1(e)\rt)\,p_1^b(e)=0,\quad\text{where}\quad e^{\theta^{a} \tau^{a} / 2} \tau^{a} e^{-\theta^{a} \tau^{a} / 2}=\Lambda_{b}^{a} (\vec{\theta} ) \tau^{b}
\ee 
is satisfied by the ansatz at all $\t$. Therefore when we find a solution (as shown below) by inserting the ansatz to Eqs.\Ref{eoms1} and \Ref{eoms2}, the effective dynamics implied by the solution is gauge invariant since the Gauss constraint is preserved by the dynamics.

Computing derivatives of ${\bf H}_\Delta$ leads to the following result:

\begin{Theorem}

The time evolution equations \Ref{homeom1} are equivalent to the following Hamiltonian equations
\be
\frac{\mathrm{d} \theta_I}{\mathrm{d} \tau}=-\frac{\kappa}{a^{2}}\frac{\partial}{\partial p_I}H_\Delta(\vec{\theta},\vec{p}),\quad \frac{\mathrm{d} p_I}{\mathrm{d} \tau}=\frac{\kappa}{a^{2}}\frac{\partial}{\partial \theta_I}H_\Delta(\vec{\theta},\vec{p}),\label{thm1}
\ee
where $H_\Delta(\vec{\theta},\vec{p})=-C^\Delta_v|_{\delta \theta=\delta p=0}$:
\be
H_\Delta(\vec{\theta},\vec{p})&=&\frac{16 a^3 }{3 \kappa\b^{1/2}\Delta }\sqrt{  p_1 p_2 p_3} \Bigg[ \sin \left(\frac{\theta _2
   \lambda  p_2}{\sqrt{p_1 p_2 p_3}}\right) \sin \left(\frac{\theta _3 \lambda  p_3}{\sqrt{p_1
   p_2 p_3}}\right)+ \sin \left(\frac{\theta _1 \lambda  p_1}{\sqrt{p_1 p_2 p_3}}\right)
   \sin \left(\frac{\theta _2 \lambda  p_2}{\sqrt{p_1 p_2 p_3}}\right)\nonumber\\
   &&+\sin \left(\frac{\theta _1 \lambda  p_1}{\sqrt{p_1 p_2 p_3}}\right)\sin
   \left(\frac{\theta _3 \lambda  p_3}{\sqrt{p_1 p_2 p_3}}\right)-\frac{3}{16} \beta^2  \Delta  \Lambda\Bigg],\label{thm2}
\ee
while Eq.\Ref{homeom2} are satisfied automatically.

\end{Theorem}

\textbf{Proof:} For a short-hand notation, ${\xi}^A=(\delta p^a_{\|}(e), \delta p^a_{\perp}(e),\delta\theta_{\|}(e),\delta\theta_{\perp}(e))$ denotes a vector of all perturbations. We have the following expansion
\be
{\bf H}=\sum_v\sqrt{\lt|\lt[C_{v}^\Delta\big|_0+\frac{\partial C_{v}^\Delta}{\partial \xi^A}\Big|_0\xi^A+O(\xi^2)\rt]^2-\frac{\a}{4}\lt[C^\Delta_{j,v}\big|_0+\frac{\partial C_{j,v}^\Delta}{\partial \xi^A}\Big|_0\xi^A+O(\xi^2)\rt]^2\rt|}
\ee
where $|_0$ means evaluation at $\xi^A=0$. The contributions to ${\partial C_{v}^\Delta}/{\partial \xi^A}|_0$ only come from the Euclidean Hamiltonian and cosmological constant terms in $C_{v}^\Delta$. ${}^3\mathcal{R}_v$ does not contribute because ${}^3\mathcal{R}_v=O(\xi^2)$. Moreover $C^\Delta_{j,v}|_0=0$ so the diffeomorphism constraint has no contribution to the linear order in $\xi^A$, and the result is independent of $\a$. We expand ${\bf H}$ and ignore $O(\xi^2)$
\be
{\bf H}&=&\sum_v\sqrt{(C_{v}^\Delta)^2\big|_0+2 C_{v}^\Delta\big|_0\frac{\partial C_{v}^\Delta}{\partial \xi^A}\Big|_0\xi^A+O(\xi^2)}\nonumber\\
&=&\sum_v\sqrt{(C_{v}^\Delta)^2\big|_0}+\sum_v\sgn\lt(C_{v}^\Delta\big|_0\rt)\frac{\partial C_{v}^\Delta}{\partial \xi^A}\Big|_0\xi^A+O(\xi^2)\nonumber\\
&=&-\sum_vC_{v}^\Delta\big|_0-\sum_v\frac{\partial C_{v}^\Delta}{\partial \xi^A}\Big|_0\xi^A+O(\xi^2)
\ee
where in the last step we use $C_v<0$ resulting from the physical dust (see Appendix \ref{BK}).

$\sum_v{\partial C_{v}^\Delta}/{\partial \xi^A}|_0$ can be obtained by a straight-forward Mathematica computation (the Mathematica code can be downloaded at \cite{github}):
\be
&&\sum_v\frac{\partial C_{v}^\Delta}{\partial \delta p^a_\perp(e_I(v))}\Big|_0=\sum_v\frac{\partial C_{v}^\Delta}{\partial \delta \theta^a_\perp(e_I(v))}\Big|_0=0,\label{pf1}\\
&&\sum_v\frac{\partial C_{v}^\Delta}{\partial \delta p^a_{\|}(e_I(v))}\Big|_0=-\frac{\partial H_\Delta(\vec{\theta},\vec{p})}{\partial p_I},\quad \sum_v\frac{\partial C_{v}^\Delta}{\partial \delta \theta^a_{\|}(e_I(v))}\Big|_0=-\frac{\partial H_\Delta(\vec{\theta},\vec{p})}{\partial \theta_I},\label{pf2}
\ee 
where $H_\Delta(\vec{\theta},\vec{p})$ is given by Eq.\Ref{thm2}. Eq.\Ref{pf1} implies constraint equations\Ref{homeom2} are automatically satisfied, while Eq.\Ref{pf2} implies Eq.\Ref{thm1}.\\
$\Box$

To contact the effective dynamics in LQC Bianchi-I model \cite{Ashtekar:2009vc}, we define 
\be
\bar{\mu}_{1}=\lambda \sqrt{\frac{p_{1}}{p_{2} p_{3}}}, \quad \bar{\mu}_{2}=\lambda \sqrt{\frac{p_{2}}{p_{1} p_{3}}},  \quad \bar{\mu}_{3}=\lambda \sqrt{\frac{p_{3}}{p_{1} p_{2}}},\quad \l=\frac{\sqrt{\Delta}}{\b^{1/2}a}. 
\ee
In terms of the conventional variables $C_I,P_I$ used in LQC,
\be
&&\theta_I=C_I,\quad p_I=\frac{P_I}{\b a^2},\\
&&\bar{\mu}_{1}=\sqrt{\Delta} \sqrt{\frac{P_{1}}{P_{2} P_{3}}}, \quad \bar{\mu}_{2}=\sqrt{\Delta}\sqrt{\frac{P_{2}}{P_{1} P_{3}}}, \quad \bar{\mu}_{3}=\sqrt{\Delta} \sqrt{\frac{P_{3}}{P_{1} P_{2}}}.
\ee
Then $H_\Delta(\vec{\theta},\vec{p})$ is reduces to the effective Hamiltonian in LQC up to an overall minus sign:
\be
H_\Delta(\vec{\theta},\vec{p})&=&\frac{16 a }{3 \b^{3/2}\kappa }\frac{1}{\left(p_{1} p_{2} p_{3}\right)^{1 / 2}}\left(\frac{\sin \left(\bar{\mu}_{1} \theta_{1}\right)}{\bar{\mu}_{1}} \frac{\sin \left(\bar{\mu}_{2} \theta_{2}\right)}{\bar{\mu}_{2}} p_{1} p_{2}+\text { cyclic terms }\right)\nonumber\\
&&-\frac{a^3 \b^{3/2}}{ \kappa}  \Lambda\left(p_{1} p_{2} p_{3}\right)^{1 / 2}\\
&=&\frac{16 }{3 \b^{2}\kappa }\frac{1}{\left(P_{1} P_{2} P_{3}\right)^{1 / 2}}\left(\frac{\sin \left(\bar{\mu}_{1} C_{1}\right)}{\bar{\mu}_{1}} \frac{\sin \left(\bar{\mu}_{2} C_{2}\right)}{\bar{\mu}_{2}} P_{1} P_{2}+\text { cyclic terms }\right)\nonumber\\
&&-\frac{\Lambda}{ \kappa}  \left(P_{1} P_{2} P_{3}\right)^{1 / 2}
\ee
The overall minus sign is due to the choice of physical dust and flowing $\t$ backward, corresponding to a negative lapse (see Appendix \ref{BK} for details).

\section{Cosmic Bounce in $\bar{\mu}$ - Scheme}\label{Cosmic Bounce}

We simplify to the homogenous and isotropy cosmology by identifying $p_I=p$ and $\theta_I=\theta$ for $I=1,2,3$. Eqs.\Ref{thm1} and \Ref{thm2} reduce to
\be
&&\frac{\mathrm{d} \theta}{\mathrm{d} \tau}=-\frac{\kappa}{a^{2}}\frac{\partial}{\partial p}H_\Delta({\theta},{p}),\quad \frac{\mathrm{d} p}{\mathrm{d} \tau}=\frac{\kappa}{a^{2}}\frac{\partial}{\partial \theta}H_\Delta({\theta},{p}),\label{iso}\\
&&H_\Delta({\theta},{p})=\frac{1}{3}H_\Delta(\vec{\theta},\vec{p})\Big|_{p_I=p,\theta_I=\theta}=\frac{16  a^3 }{3  \beta^{1/2} \kappa \Delta}\sqrt{  p^3} \lt[\sin^2\left(\frac{  \lambda }{\sqrt{p}}\theta\right)-\frac{\b^2\Delta\L}{16} \rt]
\ee 
It is convenient to make the following change of variables:
\be
V=(p a^2\b)^{3/2},\quad b=\frac{  \lambda }{\sqrt{p}}\theta.
\ee
$H_\Delta$ reduces to the LQC Hamiltonian up to a overall minus sign:
\be
H_\Delta=\frac{16}{3\b^2\kappa\Delta}V\sin^2(b)-\frac{\L}{3\kappa}V.
\ee

In terms of $(V,b)$, the evolution equations become
\be
\frac{\rmd V}{\rmd \t}=\frac{8 V \sin (2 b)}{\b\sqrt{\Delta }},\quad \frac{\rmd b}{\rmd\t}=\frac{\beta^2  \Delta  \Lambda +8 \cos (2 b)-8}{2 \b \sqrt{\Delta }}.
\ee
$H_\Delta$ is conserved in the time evolution:
\be
\frac{16}{3\b^2\kappa\Delta}V\sin^2(b)-\frac{\L}{3\kappa}V=\frac{\ce}{3}=\frac{\rho V}{3}
\ee
where $\rho$ is the energy of the physical dust (see Eq.\Ref{C2P}). This conservation law can be used to solve for $\sin(b)$, whose solution can be insert into the first evolution equation. The resulting equation in terms of the scale factor $\fa=V^{1/3}$ gives the modified Fredmann equation
\be
\lt(\frac{\dot{\fa}}{\fa}\rt)^2=\frac{16}{9} \lt[ \Lambda  \lt(1-\frac{\beta^2  \Delta  \Lambda}{16} \rt)+  \kappa  \rho  \lt(1-\frac{\beta^2  \Delta  \Lambda }{8}\rt)-\frac{\beta^2  \Delta  \kappa ^2 }{16} \rho ^2\rt]
\ee
where $\dot{\fa}=\rmd\fa/\rmd\t$ or $\rmd\fa/\rmd(-\t)$. The modified Fredmann equation reduces to the standard Fredmann equation (up to rescaling $\t$) at low density $\rho\ll1$, with the renormalized gravitational constant $\bar{\kappa}$ and cosmological constant $\bar{\L}$:
\be
\bar{\kappa}=\kappa \lt(1-\frac{\beta^2  \Delta  \Lambda }{8}\rt),\quad \bar{\L}=\Lambda  \lt(1-\frac{\beta^2  \Delta  \Lambda}{16} \rt). 
\ee
The backward time evolution stop at $\dot{\fa}=0$ and gives the Planckian critical density
\be
\rho_c=\frac{16-\beta^2  \Delta  \Lambda }{\beta^2  \Delta  \kappa }.
\ee
which is a nonzero constant independent of the conserved quantity $\ce$, in contrast to the $\mu_0$-scheme effective dynamics obtained in \cite{Han:2019vpw}. Nonzero $\rho_c$ at $\dot{\fa}=0$ indicates that the big-bang singularity is resolved and replaced by a bounce.

\section{Lattice Independence}\label{Lattice Independence}

In this section we focus on homogeneous variables Eq.\Ref{homo} and study the behavior of ${\bf H}_\Delta$ by refining the cubic lattice $\g$. 

We define a sequence of cubic lattices $\g_r$, $r=0,1,2,\cdots$, with $\g_0=\g$. If edges in $\g$ have unit coordinate lengths, every edge in $\g_r$ has a coordinate length $r^{-1}$ in the same coordinate system. The continuum limit is given by $r\to\infty$.  

A key observation in the improved Hamiltonian ${\bf H}_{\Delta}$ is that the loop holonomy $h(\Delta)$ is invariant under lattice refinement. By definition $h(\Delta)$ is a holonomy around a surface with fixed area, thus is independent of the lattice size. More specifically at homogeneous variables, we have on $\g_r$,
\be
\theta_I^{(r)}=r^{-1} \theta_I,\quad p_I^{(r)}=r^{-2} p_I, \quad \bar{\mu}_I^{(r)}=r\bar{\mu}_I,\label{scaling1}
\ee
where the superscript $(r)$ labels quantities on $\g_r$. $h_\Delta(\fs)$ in Eqs.\Ref{hDelta1} - \Ref{hDelta4} are indeed invariant under the lattice refinement. Consequently, $\sin \left(\frac{\theta _I \lambda  p_I}{\sqrt{p_I p_{I+1} p_{I+2}}}\right)$ in Eq.\Ref{thm2} doesn't scale under the refinement. Therefore $C_v^{\Delta}$ scales the same as volume
\be
C_v^{\Delta}{}^{(r)}=r^{-3}C_v^{\Delta}.\label{scaling2}
\ee
$C_v^{\Delta}{}^{(r)}$ is constant at all $v$ by homogeneity while the number of vertices scales as $|V(\g_r)|=r^3|V(\g)|$. We obtain the invariance of the Hamiltonian
\be
{\bf H}_{\Delta}^{(r)}={\bf H}_{\Delta}.\label{Hr=H}
\ee
Therefore the continuum limit of ${\bf H}_{\Delta}$ is trivial at homogeneous variables. Evaluating ${\bf H}_{\Delta}$ on any cubic lattice is equivalent to the evaluation on the continuum. 

Because $H_\Delta$ in Eqs \Ref{thm1} and \Ref{iso} equal to $-C^\Delta_v$ evaluated at homogeneous variables, the scalings \Ref{scaling1} and \Ref{scaling2} imply that Eqs.\Ref{thm1} and \Ref{iso} are invariant under the lattice refinement. Therefore the cosmological effective dynamics, the predictions of bounce and critical density are independent of the lattice refinement, so can be understood as results at the continuum limit.

If there exists an operator $\hat{\bf H}_{\Delta}{}^{(r)}$ such that ${\bf H}_{\Delta}^{(r)}=\langle\psi^{(r)}|\hat{\bf H}_{\Delta}{}^{(r)}|\psi^{(r)}\rangle$ with a sequence of states $\psi^{(r)}$ representing the homogeneous spatial geometry in the LQG Hilbert space on $\g_r$, Eq.\Ref{Hr=H} becomes 
\be
\langle\psi^{(r)}|\hat{\bf H}_{\Delta}{}^{(r)}|\psi^{(r)}\rangle=\langle\psi^{(0)}|\hat{\bf H}_{\Delta}{}^{(0)}|\psi^{(0)}\rangle, \label{RG}
\ee
which suggests ${\bf H}_{\Delta}$ at homogeneous variables is a fix point in the Hamiltonian renormalization proposed in \cite{Lang:2017beo}. However, constructions of the operator and states such as $\hat{\bf H}_{\Delta}$ and $\psi^{(r)}$ are beyond the scope of this paper. 

Interestingly, from the viewpoint of lattice field theory, the trivially of the lattice refinement Eq.\Ref{Hr=H} suggests that the theory is scaling invariant in 3 dimensions at the state of homogeneous spatial geometry, and is conformal invariant at the state of homogeneous and isotropic spatial geometry\footnote{Scaling, translational, and rotational invariance imply conformal invariance.}. This 3-dimensional conformal invariance may relate to the (anti)-de Sitter/conformal field theory correspondence.

\section{Outlook}

In this section, we discuss a few interesting perspectives which have not yet been addressed in this paper, but will be studied and reported in the future. 

Firstly as has been mentioned a few times in the above discussion, it is useful to develop an operator formalism for ${\bf H}_\Delta$ and find a series of semiclassical states $\psi^{(r)}$ to realize Eq.\Ref{RG}, in particular from the perspective of the Hamiltonian renormalization. We should also search for a quantum derivation of $S_{eff}$ from the path integral formula in \cite{Han:2019vpw}.

Secondly, a research currently undergoing is to generalize the effective dynamics from homogeneous cosmology to other spacetimes, since the effective equations we obtained in Section \ref{Path Integral and Effective Equations} are for the full theory. We have applied these equations to study e.g. cosmological perturbations and spherical symmetric black holes (in both Kantowski-Sachs foliation and global Kruskal foliation). This direction has two strategies:

\begin{itemize}

\item Similar to the present strategy, we may implement ansatz that respects the symmetry of the expected solution, then simplify and solve the effective equations \cite{prep1,prep2}. It is also similar to the strategy of analytically solving Einstein equation to obtain e.g. black holes and cosmology.

\item A different strategy is similar to Numerical Relativity. We are developing a numerical code implementing the effective equations \Ref{eoms1} - \Ref{eoms2} of the full theory. Eqs.\Ref{eoms1} and \Ref{eoms2} can be cast into a formulation similar to the evolution equations used in Numerical Relativity, thus standard numerical method such as 4th order runge kutta can be applied to our effective equations. Numerical solutions can be generated by specifying suitable initial conditions of $p^a(e),\theta^a(e)$. As an initial application of the numerical code, we find that the cosmological solution is unique provided the homogeneous and isotropic initial data \cite{Han:2019unique}. The path integral in \cite{Han:2019vpw} has a unique critical point when initial and final cosmological coherent states can be related by effective equations (so has an oscillatory behavior as $t\to0$), or has no critical points if they are not related by effective equations (so exponentially suppressed as $t\to0$). 

\end{itemize}

Lastly, the analysis of this paper focus on improving the Hamiltonian whose Lorentzian part is the scalar curvature as in \cite{Alesci:2014aza,Assanioussi:2015gka}. The next step may be the generalization to Thiemann's Lorentzian Hamiltonian which involving $K$ as the commutator between Euclidean Hamiltonian and volume.


\section*{Acknowledgements}

This work receives support from the National Science Foundation through grant PHY-1912278. 


\appendix

\section{Physical and Phantom Brown-Kucha\v{r} Dust}\label{BK}

We denote by $S_{BKD}$ the dust actions of Brown-Kucha\v{r} model \cite{Brown:1994py,Kuchar:1990vy,Giesel:2007wn,Giesel:2012rb}:
\be
S_{BKD}[\rho,g_{\mu\nu},T,S^j,W_j]&=& -\frac{1}{2}\int\rmd^4x\ \sqrt{|\det(g)|}\ \rho\ [g^{\mu\nu}U_\mu U_\nu+1],\label{dustaction}\\
U_\mu&=&-\partial_\mu T+W_j\partial_\mu S^j,
\ee
where scalars $T, S^{j=1,2,3}$ form the dust reference frame, and $\rho,\ W_j$ are Lagrangian multipliers. $\rho$ is interpreted as the dust energy density. When we couple $S_{BKD}$ to Einstein gravity and carry out the Hamiltonian analysis \cite{Giesel:2012rb}, we obtain following constraints:
\be
C^{tot}&=&C+\frac{1}{2}\left[\frac{P^{2} / \rho}{\sqrt{\operatorname{det}(q)}}+\sqrt{\operatorname{det}(q)} \rho\left(q^{\a \b} U_{\a} U_{\b}+1\right)\right]=0,\label{C}\\
C^{tot}_\a&=&C_\a+PT_{,\a}-P_jS^j_{,\a}=0,\label{Ca}\\
\rho^2&=&\frac{P^2}{\det(q)}\lt(1+q^{\a\b}U_\a U_\b\rt)^{-1},\label{rhoP}\\
W_j&=&P_j/P,\label{WP}
\ee
where $\a,\b$ are spatial coordinate index, $P,P_j$ are momenta conjugate to $T,S^j$, and $C,C_\a$ are Hamiltonian and diffeomorphism constraints of gravity. Firstly Eq.\Ref{rhoP} can be solved by
\be
\rho=\eps\frac{P}{\sqrt{\det(q)}}\lt(1+q^{\a\b}U_\a U_\b\rt)^{-1/2}, \quad \eps=\pm1.
\ee
The sign ambiguity $\eps$ may be fixed to $\eps=1$ by physical requirement that $U$ is timelike and future pointing \cite{Giesel:2007wi}, so that $\sgn(P)=\sgn(\rho)$. Inserting this solution to Eq.\Ref{C} and using Eq.\Ref{WP} lead to
\be
C=-P\sqrt{1+q^{\a\b}C_\a C_\b/P^2}.
\ee
Thus $-\sgn(C)=\sgn(P)=\sgn(\rho)$. When we consider dust coupling to pure gravity, we must have the physical dust $\rho,P>0$ to fulfill the energy condition as in \cite{Brown:1994py}. If we consider to couple additional matter fields to make $C>0$, we can let $\rho,P<0$ which corresponds to the phantom dust as in \cite{Giesel:2007wn,Giesel:2007wi}. The case of phantom dust may not violate the usual energy condition due to the presence of additional matter fields. We can solve $P,P_j$ from Eqs.\Ref{C} and \Ref{Ca}
\be
&&P=-h, \quad h=\begin{cases}-\sqrt{C^2-q^{\a\b}C_\a C_\b}&\  \text{physical dust},\\
\sqrt{C^2-q^{\a \b}C_\a C_\b} &\ \text{phantom dust},
\end{cases}\label{P=-h}\\
&&P_j=-S^\a_j\lt(C_\a-hT_{,\a}\rt)
\ee
which are strongly Poisson commutative constraints. $S^\a_j$ is the inverse matrix of $\partial_\a S^j$ ($\a=1,2,3$). In deriving above constraints, we find at an intermediate step that $P^2 = C^2- q^{\a\b}C_\a C_\b$. Hence, while the argument of the square root is not manifestly positive, it is constrained to be positive. The physical dust requires $C<0$ while the phantom dust requires $C>0$.

Gauge invariant Dirac observables are constructed relationally by parametrizing gravity canonical variables $(A,E)$ with values of dust fields $T(x)\equiv\t,S^j(x)\equiv\sig^j$, i.e. $A_j^a(\sig,\t)=A_j^a(x)|_{T(x)\equiv\t,\,S^j(x)\equiv\sig^j}$ and $E^j_a(\sig,\t)=E^j_a(x)|_{T(x)\equiv\t,\,S^j(x)\equiv\sig^j}$, where $\sig,\t$ are physical space and time coordinates in the dust frame. Here $j=1,2,3$ is the coordinate index of the dust frame (e.g. $A_j=A_\a S^\a_j$), and $a=1,2,3$ is the su(2) index.

Both $A_j^a(\sig,\t)$ and $E^j_a(\sig,\t)$ are Dirac observables. They satisfy the standard Poisson bracket $\{E^i_a(\sig,\t),A_j^b(\sig',\t)\}=\frac{1}{2}\kappa \b\delta^{i}_j\delta^b_a\delta^{3}(\sig,\sig')$ where $\b$ is the Barbero-Immirzi parameter and $\kappa=16\pi G$. The phase space $\cp$ of $A_j^a(\sig,\t),E^j_a(\sig,\t)$ is free of Hamiltonian and diffeomorphism constraints, and all phase space functions are Dirac observables.

The evolution in physical time $\t$ is generated by the physical Hamiltonian ${\bf H}$ given by integrating $h$ on the constant $T=\t$ slice $\cs$ (The constant $\t$ slice $\cs$ is coordinated by the value of dust scalars $S^j=\sig^j$ thus is often referred to as the dust space \cite{Giesel:2007wn,Giesel:2012rb}). From Eq.\Ref{P=-h}, we find that ${\bf H}$ is negative for physical dust while is positive for phantom dust. We flip the direction of the time flow $\t\to -\t$ thus ${\bf H} \to -{\bf H}$ for physical dust so that we define a positive Hamiltonian in both cases:
\be
\mathbf{H}=\int_\cs\rmd^3\sig\, \sqrt{C(\sig,\t)^2-\frac{1}{4}\sum_{j=1}^3C_j(\sig,\t)C_j(\sig,\t)}.\label{ham1}
\ee 
Here $C$ and $C_j=e_j^\a C_\a$ are parametrized in the dust frame. In terms of $A_j^a(\sig,\t)$ and $E^j_a(\sig,\t)$:
\be
C&=&-\frac{2}{\kappa\sqrt{\det(q)}}\tr\lt(F_{jk}\lt[E^j,E^k\rt]\rt)+\frac{2({1-s\b^2})}{\kappa\sqrt{\det(q)}}\tr\lt(\lt[K_j,K_k\rt]\lt[E^j,E^k\rt]\rt),\label{Csigmatau}\\
C_j&=&-\frac{2}{\kappa\sqrt{\det(q)}}\tr\lt(\t_j F_{kl}\lt[E^k,E^l\rt]\rt)\label{Cjsigmatau}.
\ee  
where $E^j=E^j_a\t^a/2$, the extrinsic curvature $K_j=K_j^a\t^a/2$, and $F_{jk}=F_{jk}^a\t^a/2$ is the curvature of $A_j=A_j^a\t^a/2$. $\t^a=-i(\text{Pauli matrix})^a$. $s=1$ or $-1$ corresponds respectively to the Euclidean or Lorentzian signature. The physical Hamiltonian $\mathbf{H}$ generates the $\t$-time evolution:
\be
\frac{\rmd f}{\rmd\t}=\lt\{\mathbf{H}, f\rt\},
\ee
for all phase space function $f$ of $A_j^a(\sig,\t)$ and $E^j_a(\sig,\t)$. 

The gravity-dust models only deparametrize the Hamiltonian and diffeomorphism constraints, while the SU(2) Gauss constraint $D_j E^j=0$ still has to be imposed to the classical phase space. In addition, we impose some non-holonomic constraints to the phase space: $C(\sig,\t)^2-\frac{1}{4}\sum_{j=1}^3C_j(\sig,\t)C_j(\sig,\t)\geq 0$ and $C<0$ for physical dust ($C>0$ for phantom dust).

The variation of ${\bf H}$ is given by
\be
\delta{\bf H}=\int_\cs\rmd^3\sig \lt(\frac{C}{|h|}\delta C -q^{\a\b}\frac{C_\b}{|h|}\delta C_\a+\frac{1}{2|h|}q^{\a\g}q^{\b\rho}C_\g C_\rho\delta q_{\a\b}\rt),
\ee
where ${C}/{|h|}$ is negative (positive) for physical (phantom) dust, and the last term vanishes in the case of spatial-homogeneous solution. If we compare $\delta{\bf H}$ to the variation of Hamiltonian $H_{GR}$ of pure gravity in the absence of dust
\be
\delta H_{GR}=\int_\cs\rmd^3\sig \lt(N \delta C +N^\a \delta C_\a\rt),
\ee
where the lapse and shift $N,N^\a$ are constant Lagrangian multipliers, we find $\delta{\bf H}$ and $\delta H_{GR}$ coincide at the spatial-homogeneous solution, provided we identify 
\be
N=\frac{C}{|h|}, \quad N^\a=-q^{\a\b}\frac{C_\b}{|h|}.
\ee
Therefore $N$ is negative (positive) for the physical (phantom) dust. Negative $N$ for the physical dust relates to the flip $\t\to-\t$ for making Hamiltonian positive.

In the case of spatial-homogeneous solution, $C_\a=0$ and $U_\a=C_\a/P=0$ so that the Hamiltonian density
\be
\sqrt{C^2}=\begin{cases} 
P =\rho\sqrt{\det(q)}& \text{for physical dust}\ \rho>0,\\
-P =-\rho\sqrt{\det(q)}& \text{for phantom dust}\ \rho<0,
\end{cases}\label{C2P}
\ee
which is a conserved quantity in the effective dynamics.

\bibliographystyle{jhep}

\bibliography{muxin}

\providecommand{\href}[2]{#2}\begingroup\raggedright\begin{thebibliography}{10}

\bibitem{book}
T.~Thiemann, {\em Modern Canonical Quantum General Relativity}.
\newblock Cambridge University Press, 2007.

\bibitem{review}
M.~Han, W.~Huang, and Y.~Ma, {\it {Fundamental structure of loop quantum
  gravity}},  {\em Int.J.Mod.Phys.} {\bf D16} (2007) 1397--1474,
  [\href{http://arxiv.org/abs/gr-qc/0509064}{{\tt gr-qc/0509064}}].

\bibitem{Bojowald:2001xe}
M.~Bojowald, {\it {Absence of singularity in loop quantum cosmology}},  {\em
  Phys. Rev. Lett.} {\bf 86} (2001) 5227--5230,
  [\href{http://arxiv.org/abs/gr-qc/0102069}{{\tt gr-qc/0102069}}].

\bibitem{Ashtekar:2006wn}
A.~Ashtekar, T.~Pawlowski, and P.~Singh, {\it {Quantum Nature of the Big Bang:
  Improved dynamics}},  {\em Phys. Rev.} {\bf D74} (2006) 084003,
  [\href{http://arxiv.org/abs/gr-qc/0607039}{{\tt gr-qc/0607039}}].

\bibitem{Singh:2009mz}
P.~Singh, {\it {Are loop quantum cosmos never singular?}},  {\em Class. Quant.
  Grav.} {\bf 26} (2009) 125005, [\href{http://arxiv.org/abs/0901.2750}{{\tt
  arXiv:0901.2750}}].

\bibitem{Assanioussi:2019iye}
M.~Assanioussi, A.~Dapor, K.~Liegener, and T.~Pawlowski, {\it {Emergent de
  Sitter epoch of the Loop Quantum Cosmos: a detailed analysis}},
  \href{http://arxiv.org/abs/1906.05315}{{\tt arXiv:1906.05315}}.

\bibitem{Ashtekar:2018cay}
A.~Ashtekar, J.~Olmedo, and P.~Singh, {\it {Quantum extension of the Kruskal
  spacetime}},  {\em Phys. Rev.} {\bf D98} (2018), no.~12 126003,
  [\href{http://arxiv.org/abs/1806.02406}{{\tt arXiv:1806.02406}}].

\bibitem{Ashtekar:2018lag}
A.~Ashtekar, J.~Olmedo, and P.~Singh, {\it {Quantum Transfiguration of Kruskal
  Black Holes}},  {\em Phys. Rev. Lett.} {\bf 121} (2018), no.~24 241301,
  [\href{http://arxiv.org/abs/1806.00648}{{\tt arXiv:1806.00648}}].

\bibitem{Assanioussi:2019twp}
M.~Assanioussi, A.~Dapor, and K.~Liegener, {\it {Perspectives on the dynamics
  in loop effective black hole interior}},
  \href{http://arxiv.org/abs/1908.05756}{{\tt arXiv:1908.05756}}.

\bibitem{Gambini:2013hna}
R.~Gambini, J.~Olmedo, and J.~Pullin, {\it {Quantum black holes in Loop Quantum
  Gravity}},  {\em Class. Quant. Grav.} {\bf 31} (2014) 095009,
  [\href{http://arxiv.org/abs/1310.5996}{{\tt arXiv:1310.5996}}].

\bibitem{BenAchour:2018khr}
J.~Ben~Achour, F.~Lamy, H.~Liu, and K.~Noui, {\it {Polymer Schwarzschild black
  hole: An effective metric}},  {\em EPL} {\bf 123} (2018), no.~2 20006,
  [\href{http://arxiv.org/abs/1803.01152}{{\tt arXiv:1803.01152}}].

\bibitem{Rovelli:2014cta}
C.~Rovelli and F.~Vidotto, {\it {Planck stars}},  {\em Int. J. Mod. Phys.} {\bf
  D23} (2014), no.~12 1442026, [\href{http://arxiv.org/abs/1401.6562}{{\tt
  arXiv:1401.6562}}].

\bibitem{Han:2016fgh}
M.~Han and M.~Zhang, {\it {Spinfoams near a classical curvature singularity}},
  {\em Phys. Rev.} {\bf D94} (2016), no.~10 104075,
  [\href{http://arxiv.org/abs/1606.02826}{{\tt arXiv:1606.02826}}].

\bibitem{Han:2019vpw}
M.~Han and H.~Liu, {\it {Effective Dynamics from Coherent State Path Integral
  of Full Loop Quantum Gravity}},  \href{http://arxiv.org/abs/1910.03763}{{\tt
  arXiv:1910.03763}}.

\bibitem{Bojowald:2006da}
M.~Bojowald, {\it {Loop quantum cosmology}},  {\em Living Rev. Rel.} {\bf 8}
  (2005) 11, [\href{http://arxiv.org/abs/gr-qc/0601085}{{\tt gr-qc/0601085}}].

\bibitem{Ashtekar:2008zu}
A.~Ashtekar, {\it {Loop Quantum Cosmology: An Overview}},  {\em Gen. Rel.
  Grav.} {\bf 41} (2009) 707--741, [\href{http://arxiv.org/abs/0812.0177}{{\tt
  arXiv:0812.0177}}].

\bibitem{Agullo:2016tjh}
I.~Agullo and P.~Singh, {\it {Loop Quantum Cosmology}},  in {\em Loop Quantum
  Gravity: The First 30 Years} (A.~Ashtekar and J.~Pullin, eds.), pp.~183--240.
\newblock WSP, 2017.
\newblock \href{http://arxiv.org/abs/1612.01236}{{\tt arXiv:1612.01236}}.

\bibitem{Taveras:2008ke}
V.~Taveras, {\it {Corrections to the Friedmann Equations from LQG for a
  Universe with a Free Scalar Field}},  {\em Phys. Rev.} {\bf D78} (2008)
  064072, [\href{http://arxiv.org/abs/0807.3325}{{\tt arXiv:0807.3325}}].

\bibitem{Alesci:2013xd}
E.~Alesci and F.~Cianfrani, {\it {Quantum-Reduced Loop Gravity: Cosmology}},
  {\em Phys. Rev.} {\bf D87} (2013), no.~8 083521,
  [\href{http://arxiv.org/abs/1301.2245}{{\tt arXiv:1301.2245}}].

\bibitem{Bodendorfer:2014vea}
N.~Bodendorfer, {\it {Quantum reduction to Bianchi I models in loop quantum
  gravity}},  {\em Phys. Rev.} {\bf D91} (2015), no.~8 081502,
  [\href{http://arxiv.org/abs/1410.5608}{{\tt arXiv:1410.5608}}].

\bibitem{Bodendorfer:2015hwl}
N.~Bodendorfer, {\it {An embedding of loop quantum cosmology in $(b,v)$
  variables into a full theory context}},  {\em Class. Quant. Grav.} {\bf 33}
  (2016), no.~12 125014, [\href{http://arxiv.org/abs/1512.00713}{{\tt
  arXiv:1512.00713}}].

\bibitem{Alesci:2016rmn}
E.~Alesci and F.~Cianfrani, {\it {Improved regularization from Quantum Reduced
  Loop Gravity}},  \href{http://arxiv.org/abs/1604.02375}{{\tt
  arXiv:1604.02375}}.

\bibitem{Dapor:2017rwv}
A.~Dapor and K.~Liegener, {\it {Cosmological Effective Hamiltonian from full
  Loop Quantum Gravity Dynamics}},  {\em Phys. Lett.} {\bf B785} (2018)
  506--510, [\href{http://arxiv.org/abs/1706.09833}{{\tt arXiv:1706.09833}}].

\bibitem{Engle:2007zz}
J.~Engle, {\it {Relating loop quantum cosmology to loop quantum gravity:
  Symmetric sectors and embeddings}},  {\em Class. Quant. Grav.} {\bf 24}
  (2007) 5777--5802, [\href{http://arxiv.org/abs/gr-qc/0701132}{{\tt
  gr-qc/0701132}}].

\bibitem{2016arXiv160105531H}
M.~{Hanusch}, {\it {Invariant Connections and Symmetry Reduction in Loop
  Quantum Gravity}},  {\em arXiv e-prints} (Jan, 2016) arXiv:1601.05531,
  [\href{http://arxiv.org/abs/1601.05531}{{\tt arXiv:1601.05531}}].

\bibitem{Fleischhack:2010zt}
C.~Fleischhack, {\it {Loop Quantization and Symmetry: Configuration Spaces}},
  {\em Commun. Math. Phys.} {\bf 360} (2018), no.~2 481--521,
  [\href{http://arxiv.org/abs/1010.0449}{{\tt arXiv:1010.0449}}].

\bibitem{Rovelli:2008aa}
C.~Rovelli and F.~Vidotto, {\it {Stepping out of Homogeneity in Loop Quantum
  Cosmology}},  {\em Class. Quant. Grav.} {\bf 25} (2008) 225024,
  [\href{http://arxiv.org/abs/0805.4585}{{\tt arXiv:0805.4585}}].

\bibitem{Calcagni:2014tga}
G.~Calcagni, {\it {Loop quantum cosmology from group field theory}},  {\em
  Phys. Rev.} {\bf D90} (2014), no.~6 064047,
  [\href{http://arxiv.org/abs/1407.8166}{{\tt arXiv:1407.8166}}].

\bibitem{Dapor:2019mil}
A.~Dapor, K.~Liegener, and T.~Pawlowski, {\it {Challenges in Recovering a
  Consistent Cosmology from the Effective Dynamics of Loop Quantum Gravity}},
  {\em Phys. Rev.} {\bf D100} (2019), no.~10 106016,
  [\href{http://arxiv.org/abs/1910.04710}{{\tt arXiv:1910.04710}}].

\bibitem{Giesel:2007wi}
K.~Giesel, S.~Hofmann, T.~Thiemann, and O.~Winkler, {\it {Manifestly
  Gauge-Invariant General Relativistic Perturbation Theory. I. Foundations}},
  {\em Class. Quant. Grav.} {\bf 27} (2010) 055005,
  [\href{http://arxiv.org/abs/0711.0115}{{\tt arXiv:0711.0115}}].

\bibitem{Giesel:2007wn}
K.~Giesel and T.~Thiemann, {\it {Algebraic quantum gravity (AQG). IV. Reduced
  phase space quantisation of loop quantum gravity}},  {\em Class. Quant.
  Grav.} {\bf 27} (2010) 175009, [\href{http://arxiv.org/abs/0711.0119}{{\tt
  arXiv:0711.0119}}].

\bibitem{QSD}
T.~Thiemann, {\it {Quantum spin dynamics (QSD)}},  {\em Class. Quant. Grav.}
  {\bf 15} (1998) 839--873, [\href{http://arxiv.org/abs/gr-qc/9606089}{{\tt
  gr-qc/9606089}}].

\bibitem{Lang:2017beo}
T.~Lang, K.~Liegener, and T.~Thiemann, {\it {Hamiltonian renormalisation I:
  derivation from Osterwalder-Schrader reconstruction}},  {\em Class. Quant.
  Grav.} {\bf 35} (2018), no.~24 245011,
  [\href{http://arxiv.org/abs/1711.05685}{{\tt arXiv:1711.05685}}].

\bibitem{Alesci:2014aza}
E.~Alesci, M.~Assanioussi, and J.~Lewandowski, {\it {Curvature operator for
  loop quantum gravity}},  {\em Phys. Rev.} {\bf D89} (2014), no.~12 124017,
  [\href{http://arxiv.org/abs/1403.3190}{{\tt arXiv:1403.3190}}].

\bibitem{Assanioussi:2015gka}
M.~Assanioussi, J.~Lewandowski, and I.~Makinen, {\it {New scalar constraint
  operator for loop quantum gravity}},  {\em Phys. Rev.} {\bf D92} (2015),
  no.~4 044042, [\href{http://arxiv.org/abs/1506.00299}{{\tt
  arXiv:1506.00299}}].

\bibitem{github}
M.~Han and H.~Liu.
  \url{https://github.com/LQG-Florida-Atlantic-University/Hamiltonian}, 2019.

\bibitem{Ashtekar:2009vc}
A.~Ashtekar and E.~Wilson-Ewing, {\it {Loop quantum cosmology of Bianchi I
  models}},  {\em Phys. Rev.} {\bf D79} (2009) 083535,
  [\href{http://arxiv.org/abs/0903.3397}{{\tt arXiv:0903.3397}}].

\bibitem{prep1}
M.~Han, H.~Li, and H.~Liu, {\it {Cosmological perturbation theory in full loop
  quantum gravity}},  {\em in preparation}.

\bibitem{prep2}
A.~Dapor, M.~Han, and H.~Liu, {\it {Effective Dynamics of Black Holes in Full
  Loop Quantum Gravity}},  {\em in preparation}.

\bibitem{Han:2019unique}
M.~Han and H.~Liu, {\it {The uniqueness of cosmological solution in full loop
  quantum gravity}},  {\em to appear}.

\bibitem{Brown:1994py}
J.~D. Brown and K.~V. Kuchar, {\it {Dust as a standard of space and time in
  canonical quantum gravity}},  {\em Phys. Rev.} {\bf D51} (1995) 5600--5629,
  [\href{http://arxiv.org/abs/gr-qc/9409001}{{\tt gr-qc/9409001}}].

\bibitem{Kuchar:1990vy}
K.~V. Kuchar and C.~G. Torre, {\it {Gaussian reference fluid and interpretation
  of quantum geometrodynamics}},  {\em Phys. Rev.} {\bf D43} (1991) 419--441.

\bibitem{Giesel:2012rb}
K.~Giesel and T.~Thiemann, {\it {Scalar Material Reference Systems and Loop
  Quantum Gravity}},  {\em Class. Quant. Grav.} {\bf 32} (2015) 135015,
  [\href{http://arxiv.org/abs/1206.3807}{{\tt arXiv:1206.3807}}].

\end{thebibliography}\endgroup

\end{document}